# The Influence of Chemical Strains on the Electrocaloric Response, Polarization Morphology, Tetragonality and Negative Capacitance Effect of Ferroelectric Core-Shell Nanorods and Nanowires

Anna N. Morozovska[1*], Eugene A. Eliseev[2], Olha A. Kovalenko[2], and Dean R. Evans[3†]

[1] Institute of Physics, National Academy of Sciences of Ukraine,

46, pr. Nauky, 03028 Kyiv, Ukraine

[2] Frantsevich Institute for Problems in Materials Science, National Academy of Sciences of Ukraine

Omeliana Pritsaka str., 3, Kyiv, 03142, Ukraine

[3] Air Force Research Laboratory, Materials and Manufacturing Directorate, Wright-Patterson Air Force Base, Ohio, 45433, USA

**Abstract**

Using Landau-Ginzburg-Devonshire (LGD) approach we proposed the analytical description of the chemical strains influence on the spontaneous polarization and electrocaloric response in ferroelectric core-shell nanorods. We postulate that the nanorod core presents a defect-free single-crystalline ferroelectric material, and the elastic defects are accumulated in the ultra-thin shell, where they can induce tensile or compressive chemical strains. The finite element modeling (FEM) based on the LGD approach reveals transitions of domain structure morphology induced by the chemical strains in the $BaTiO_3$ nanorods. Namely, tensile chemical strains induce and support the single-domain state in the central part of the nanorod, while the curled domain structures appear near the unscreened or partially screened ends of the rod. The vortex-like domains propagate toward the central part of the rod and fill it entirely, when the rod is covered by a shell with compressive chemical strains above some critical value. The critical value depends on the nanorod sizes, aspect ratio, and screening conditions at its ends. Both analytical theory and FEM predict that the tensile chemical strains in the shell increase the nanorod polarization, lattice tetragonality, and electrocaloric response well-above the values corresponding to the bulk material. The physical reason of the increase is the strong electrostriction coupling between the mismatch-type elastic strains induced in the core by the chemical strains in the shell. Comparison with the earlier XRD data confirmed an increase of tetragonality ratio in tensiled $BaTiO_3$ nanorods compared to the bulk material. Obtained analytical expressions, which are suitable for the description of strain-induced changes in a wide class of multiaxial ferroelectric core-shell nanorods and nanowires, can be useful for strain engineering of advanced ferroelectric nanomaterials for energy storage, harvesting, electrocaloric applications and negative capacitance elements.

---

[*] corresponding author, e-mail: anna.n.morozovska@gmail.com
[†] Corresponding author: dean.evans@afrl.af.mil

## I. INTRODUCTION

The influence of shape and size effects, defects, and elastic strains on the phase state, polar and structural properties, and related working performances of various nanosized ferroelectrics is still poorly explored. In particular, the physical explanation and theoretical description of the strongly enhanced spontaneous polarization and lattice tetragonality observed in BaTiO$_3$ core-shell ferroelectric nanoparticles [1, 2, 3, 4] have been absent for a long time [5]. Recent X-ray spectroscopic measurements [2] revealed a large Ti-cation off-centering in 10-nm quasi-spherical BaTiO$_3$ core-shell nanoparticles near 300 K confirmed by the tetragonality ratio $\frac{c}{a} \approx 1.0108$, which is higher than the bulk value, $\frac{c}{a} \approx 1.010$, and significantly higher in comparison with $\frac{c}{a} \approx 1.0075$ for 50 nm nanoparticles. The off-centering of Ti-cations is a key factor in producing the enhanced spontaneous polarization (up to 130 µC/cm$^2$ at room temperature) in the core-shell nanoparticles, and the barium oleate component in the core-shell matrix (resulting from mechanochemical synthesis during the ball-milling process [6]) stabilizes the enhanced polar structural phase of the BaTiO$_3$ core. Only recently the theoretical models [7, 8, 9], which postulate the appearance of elastic strains caused by elastic defects accumulated in the shell, have been proposed, and numerical and analytical solutions for the strain-induced polarization changes in spherical core-shell nanoparticles have been derived.

Depending on the nature of elastic defects (e.g., dilatation centers, such as oxygen or cation vacancies, divacancies, OH-complexes, or isovalent impurity atoms), the defects can create compressive or tensile elastic strains in the oxide ferroelectrics, which are usually called chemical (or compositional) strains [10, 11]. Instead of "chemical" or "compositional" strains we use more narrow terminology in Refs.[7 - 9], as well as in some places in this work, namely "Vegard" strains [12, 13]. Furthermore, we consider that the strain is linearly proportional to the concentration of elastic defects (the Vegard law for chemical strains), and the proportionality coefficient is named the Vegard tensor. Assuming that the formation energy of elastic defects is much smaller near the surface than in the bulk of the ferroelectric [14], elastic defects (and corresponding strains) are accumulated in a thin layer under the surface. It was shown that the Vegard strains are responsible for the strong increase of the Curie temperature (above 440 K) and tetragonality (up to 1.032) near the surface of a BaTiO$_3$ film with injected oxygen vacancies [15].

To the best of our knowledge, analytical solutions for the strain-induced polarization changes for other shapes of core-shell ferroelectric nanoparticles are absent. However, an enhanced polarization, electrocaloric response, and high lattice tetragonality can be observed in non-spherical nanoparticles (e.g., $c/a \approx 1.013$ is observed for BaTiO$_3$ nanorods and nanowires [16]), where the core-shell structure can be formed spontaneously, because various defects are accumulated at the surface and under the surface due to the strong (e.g., exponential) lowering of the defect formation energy when approaching the surface [14]. Thus, analytical solutions are important for fundamental physics and can help to achieve significant progress in the energy storage [17, 18, 19, 20], harvesting [21] and electrocaloric applications [22] of the non-spherical ferroelectric core-shell nanoparticles.

Since the shape variation is one of the most effective means of controlling depolarization factors in ferroelectric nanoparticles, very long nanorods and nanowires with the spontaneous polarization directed along their axis have negligibly small depolarization fields, which cannot decrease the polarization. Because of this, several theoretical papers [23, 24, 25] predict the increase of a reversible spontaneous polarization in homogeneous (without the core-shell structure) ferroelectric nanorods and nanowires, when the spontaneous polarization is directed along their axis. The increase of the spontaneous polarization can appear due to the positive surface tension coefficient $\mu$ and negative electrostriction coupling coefficients $Q_{12}$ of ABO$_3$-type perovskites, because the dependence of the Curie temperature $T_C$ on the particle radius $R$ is proportional to the positive value $-\frac{4\mu}{R}Q_{12}$ in the nanowire (see e.g., Table 1 in Ref. [26]). The increase of $T_C$ becomes significant for $R \leq 5$ nm and requires very high $\mu > (5 - 10)$ N/m [26]. The flexo-chemical effect [27], being the joint action of the chemical strains and flexoelectric effect, can increase $T_C$, spontaneous polarization, and $\frac{c}{a}$ in ultra-small (5 nm or less) spherical or cylindrical BaTiO$_3$ nanoparticles, although the effect rapidly disappears with a radius increase (as $\frac{1}{R^2}$) and requires very high values of the flexoelectric coupling coefficients and strain gradients.

The chemical strains in the shell influence the core polarization almost independently on its size (until the strain relaxation via e.g., mismatch dislocations, appear). For this reason the chemical strains can significantly increase the Curie temperature, spontaneous polarization, lattice tetragonality, pyroelectric effect and electrocaloric response of the (5 - 50) nm spherical core-shell nanoparticles [7, 8, 9]. Furthermore, the higher-order electrostriction coupling [28], which needs to be considered for chemical strains higher than 1%, can be very important for a correct description of the core-shell nanoparticle polar properties [29, 30].

Using these ideas, this work analyzes polar properties of cylindrical core-shell BaTiO$_3$ nanorods and nanowires in the framework of Landau-Ginzburg-Devonshire (**LGD**) free energy functional, which includes the 8-th power of polarization, and thus allows high chemical strains in the shell and the nonlinear electrostriction coupling in the core to be considered. The analytical description of polar and electrocaloric properties of single-domain core-shell ferroelectric nanowires and long nanorods are presented in **Section II**. The finite element modeling (**FEM**) results, which show the ranges of analytical solutions applicability and demonstrate the strain-induced domain morphology in core-shell nanorods, are presented and analyzed in **Section III**. Comparison with available experiments, discussion of the controllable negative capacitance effect, and conclusion are in **Section IV**. **Supplementary Materials** [31] contain calculation details.

## II. ANALYTICAL DESCRIPTION
### A. The problem formulation for a single-domain ferroelectric nanorod

Let us consider a core-shell nanorod, whose core of radius $R_c$ and length $2L_c$ is a single-domain ferroelectric with a spontaneous polarization $\vec{P}_s$ directed along the polar axis $x_3$. The nanorod geometry is shown in **Fig. 1**. The core permittivity is $\hat{\varepsilon}_c$, which contains a background contribution, $\varepsilon_b$ [32], and the ferroelectric contribution, $\varepsilon_f$. The defect-free core is crystalline, has a tetragonal symmetry, and is insulating, since a bulk BaTiO3 has a wide band gap (around 3.4 eV). The core is covered with a crystalline shell of cubic symmetry, which has the average thickness $\Delta R = R_s - R_c$. We postulate that the shell is semiconducting due to the high concentration of free charges. The charge screening is formed spontaneously due to the multiple mechanisms of a spontaneous polarization screening by internal and external charges in nanoscale ferroelectrics (e.g., Ref. [33] and refs. therein). The free charges provide effective screening of the core spontaneous polarization and prevent domain formation. The effective screening length in the shell, λ, is relatively small (less than 1 nm), and its relative dielectric permittivity tensor, $\varepsilon_{ij}^S$, is isotropic, $\varepsilon_{ij}^S = \delta_{ij}\varepsilon_s$, and can be large enough (e.g., several hundred or higher) which correspond to the paraelectric phase.

We postulate the presence of elastic defects (e.g., oxygen or cation vacancies, divacancies, OH-complexes, or isovalent impurity atoms) in the shell and assume that they can induce strong tensile or compressive chemical strains [10, 11]. We also assume the validity of the Vegard law [12, 13] for chemical strains: the strain is linearly proportional to the concentration $\delta c$ of elastic defects, and the proportionality coefficient is the Vegard tensor, denoted as $W_{ij}^{c,s}(\vec{r})$. The Vegard tensor, whose components can be calculated from the first principles for certain cases [10], is assumed to have a cubic symmetry in average, $W_{ij}^{c,s} = \delta_{ij}w_{c,s}$, where $\delta_{ij}$ is the Kronecker-delta symbol, $w_c$ and $w_s$ are the averaged tensor magnitudes in the core (denoted by the superscript "c") and shell (denoted by the superscript "s"), respectively. According to the Vegard law, the chemical strains $u_{ij}^{c,s}$ are equal to $W_{ij}^{c,s}\delta c$. Because we assumed that the Vegard tensor has a cubic symmetry, corresponding chemical strains have the same symmetry, namely $u_{ij}^{c,s} = \delta_{ij}u_{c,s}$. As a rule, the difference $u_s - u_c$ can reach (0.5 – 3)%, and it is unlikely that it can exceed (5 – 7)% because the concentration of elastic defects cannot exceed (5 – 10) % near the surface in accordance with many experiments [14, 15].

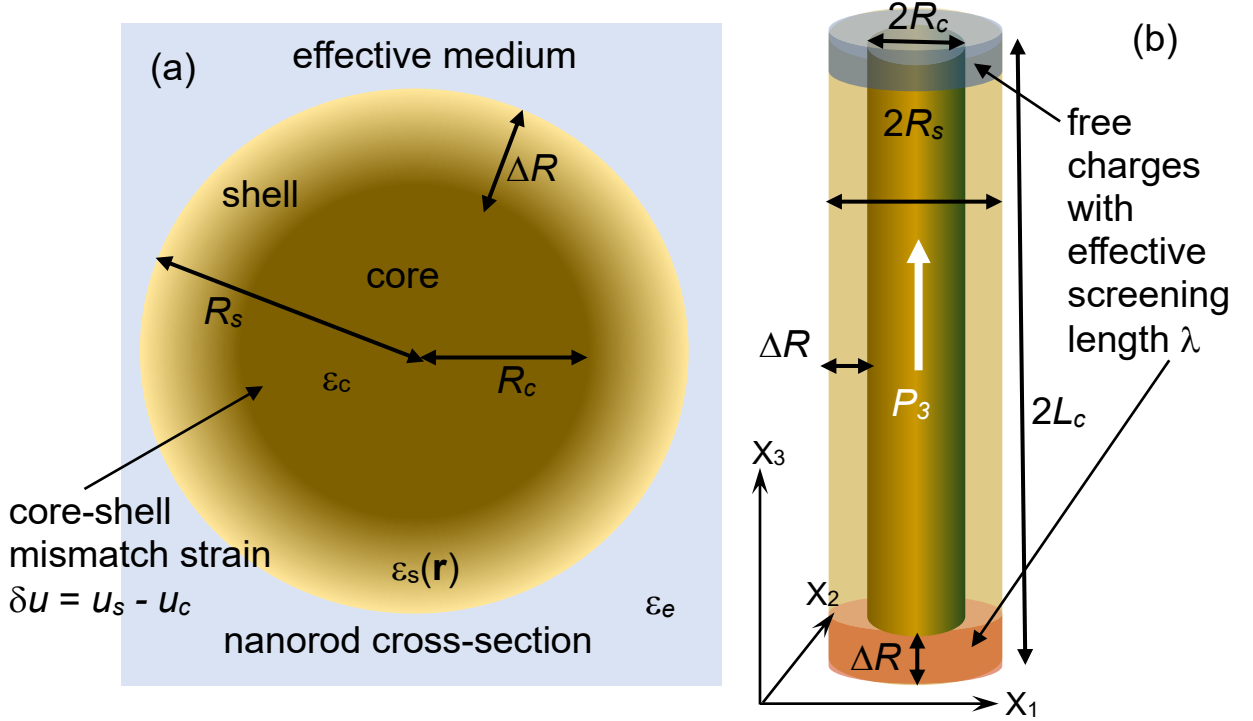

**FIGURE 1. (a)** – the radial cross-section of the core-shell ferroelectric nanorod, **(b)** – the side-view of the core-shell nanorod.

### B. Analytical solution for elastic strains, quasi-static polarization and electrocaloric properties of single-domain ferroelectric nanowires and nanorods

The LGD free energy density includes the Landau-Devonshire expansion in even powers of the polarization $P_3$ (up to the 8-th power); the Ginzburg polarization gradient energy; and the elastic, electrostriction, and flexoelectric energies, which are listed in **Table AI** in **Appendix A** [31]. Material parameters corresponding to the bulk BaTiO$_3$ are taken from Refs. [34, 35]. Components of the polarization gradient tensor are taken from Ref. [36].

In the case of natural boundary conditions for polarization vector at the ends and side surface of the nanorod (which are used hereinafter) and polarization gradient coefficients higher than $10^{-11}$ C$^{-2}$m$^3$J (which are used hereinafter), the polarization gradient effects can be neglected inside the nanowires (i.e., for $L_c \to \infty$) and long nanorods with a small effective screening length (i.e., for $R_c/L_c \ll 0.1$ and $\lambda \ll 0.1$ nm). In this case a single-domain state is revealed to have minimal energy compared to polydomain states. The field dependence of a quasi-static single-domain polarization can be found from the following equation [8]:

$$\alpha^* P_3 + \beta^* P_3^3 + \gamma P_3^5 + \delta P_3^7 = E_3^e. \tag{1}$$

The parameters $\alpha^*$, $\beta^*$, $\gamma$, and $\delta$ are the 2-nd, 4-th, 6-th, and 8-th order expansion coefficients in the $P_3$-powers of the Landau free energy. $E_3^e$ is the static external field inside the core.

The depolarization field and elastic stresses contribute to the "renormalization" of the first Landau expansion coefficient $a_1(T) = a_T(T - T_c)$, which becomes the temperature-, radius-, stress-, and screening length-dependent function $\alpha^*$ [30]:

$$\alpha^*(T, R_c, \sigma_i) = 2a_1(T) + \frac{n_d}{\varepsilon_0[\varepsilon_b n_d + \varepsilon_s(1-n_d) + \varepsilon_s n_d(L_c/\lambda)]} - 2Q_{i3}\sigma_i. \qquad (2)$$

The second term in Eq.(2) is the depolarization field contribution, which is derived in Refs.[23, 29, 30]. Here $\varepsilon_0 = 8.85$ pF/m is a universal dielectric constant, $\varepsilon_b$ is the dielectric permittivity of ferroelectric background [37]; $n_d = \frac{1}{1+(L_c/R_c)^2}$ is the "effective" depolarization factor of the nanorod in the direction of the spontaneous polarization $P_3$ [38]. The third term originates from the strain-electrostriction coupling. Here $Q_{i3}$ are the components of the second-order electrostriction tensor components and $\sigma_i$ are elastic stresses in the core, written in the Voigt notations.

Due to the nonlinear electrostriction coupling, the coefficient $\beta^*$ is "renormalized" by elastic stresses as

$$\beta^*(T, \sigma_i) = 4a_{11}(T) - 4Z_{i33}\sigma_i. \qquad (3a)$$

The values $Z_{i33}$ are the components of the higher-order electrostriction strain tensor in the Voigt notation [28]. The values

$$\gamma = 6a_{111}, \quad \delta = 8a_{1111}. \qquad (3b)$$

The temperature-dependent values $a_1(T)$ and $a_{11}(T)$ and the constants $a_{111}$ and $a_{1111}$ are listed in **Table AI** in **Appendix A** [31].

Elastic stresses and strains can be calculated analytically in a cylindrical core-shell nanorod, as derived in **Appendix B** [31]. For a very long nanorod or nanowire, the nonzero components of the core strains, $u_i^c$, written in the Voigt notations, are:

$$u_3^c = u_c + (1 - \delta V)Q_{11}P_3^2 + \delta V \delta u, \qquad (4a)$$

$$u_1^c = u_2^c = (1 - \delta V)(u_c + Q_{12}P_3^2) + \delta V \left[u_s + \frac{(s_{11}-s_{12})\delta u + (s_{11}Q_{12}-s_{12}Q_{11})P_3^2}{2(s_{11}+s_{12})}\right]. \qquad (4b)$$

Here the relative shell volume ($\delta V$) and "effective" mismatch strain ($\delta u$) are introduced as:

$$\delta V = \frac{V_s}{V} = 1 - \frac{R_c^2}{R_s^2}, \quad \delta u = u_s - u_c. \qquad (4c)$$

In Eq.(4c) we define the shell volume as $V_s = \pi L_c(R_s^2 - R_c^2)$ and the nanorod volume as $V = \pi L_c R_s^2$. In a general case, the effective mismatch strain $\delta u$ is created not only by the difference between the core and the shell chemical strains ($u_c$ and $u_s$), as assumed in Eq.(4c), but also be the lattice constants mismatch and/or different thermal expansion coefficients in the core and the shell. However, here we consider the simplest case when the elastic defects are postulated to be present in the shell only, and other contributions to $\delta u$ are absent, i.e., $\delta u \equiv u_s$ and $u_c = 0$ in Eqs.(4). Hence, below we can consider that $\delta u$ is the effective chemical strain.

Consideration of the surface tension leads to the appearance of terms proportional to $-2s_{12}\frac{\mu}{R_s}$ and $-(s_{11}+s_{12})\frac{\mu}{R_s}$ in the expressions (4a) and (4b), respectively (see **Appendix B** [31] for details). Due to the reasons discussed in the Introduction, the terms appear negligibly small (less than 0.01 %) for the considered radii of nanorods and nanowires ($R_s \geq 10$ nm) and realistic surface tension coefficient $\mu < 4$ N/m.

The tetragonality ratio of the lattice constants $c$ and $a$, is given by expression:

$$\frac{c}{a} = \frac{1+u_3^c}{1+u_1^c} \approx 1 + u_3^c - u_1^c. \tag{5a}$$

From Eq.(4), the tetragonality ratio is equal to:

$$\frac{c}{a} = 1 + (Q_{11} - Q_{12})P_3^2 - \frac{1}{2}\delta V\left[\frac{(2s_{11}+s_{12})Q_{11}-(s_{11}+2s_{12})Q_{12}}{s_{11}+s_{12}}P_3^2 + \frac{s_{11}-s_{12}}{s_{11}+s_{12}}\delta u\right]. \tag{5b}$$

The first two terms in Eq.(5b) coincide with the expression for a bulk ferroelectric with the spontaneous polarization $P_3$ in the tetragonal ferroelectric phase, which has a cubic parent phase. The next two terms, proportional to the relative shell volume $\delta V$, are caused by the elastic anisotropy between the tetragonal core and cubic shell, as well as by the effective chemical strain, $\delta u$. From Eq.(5), the non-zero tetragonality can exist in the paraelectric core-shell nanorods and is equal to $\frac{c}{a} = 1 + \frac{s_{11}-s_{12}}{s_{11}+s_{12}}\frac{\delta u}{2}\delta V$.

After substitution of the elastic strains from Eq.(4) into Eq.(1) we obtain the equation of state for the electric polarization $P_3$:

$$\alpha_R P_3 + \beta_R P_3^3 + \gamma P_3^5 + \delta P_3^7 = E_3^e. \tag{6}$$

The renormalized coefficients in Eq.(6) are given by expressions:

$$\alpha_R = 2\left\{a_1 - \delta u\, \delta V \frac{Q_{11}+Q_{12}}{s_{11}+s_{12}}\right\} + \frac{n_d}{\varepsilon_0[\varepsilon_b n_d + \varepsilon_s(1-n_d) + n_d(L_c/\lambda)]}, \tag{7a}$$

$$\beta_R = 4\left\{a_{11} + \delta V \frac{s_{11}(Q_{11}^2+Q_{12}^2)-2s_{12}Q_{11}Q_{12}}{2(s_{11}^2-s_{12}^2)}\right\} - 8\frac{Z_{211}}{s_{11}+s_{12}}\delta V \delta u. \tag{7b}$$

For a very thick shell (i.e., for $\delta V \to 1$ at $R_s \gg R_c$) expressions (7) transform into the well-known expressions [39] for the renormalized Landau coefficients in the ferroelectric thin film with in-plane spontaneous polarization, clamped to the infinitely thick substrate. In the case the "effective" strain, $\delta u$, is determined by the different chemical strains, and/or lattice constants, and/or thermal expansion coefficients in the film and its substrate. In the opposite case of a very thin shell (i.e., for $\delta V \to 0$ at $R_s \to R_c$) expressions (7) transform to the coefficients of a bulk ferroelectric.

The field dependence of a static single-domain pyroelectric coefficient $\Pi_3$ and the electrocaloric (**EC**) temperature change $\Delta T_{EC}$ in an external field $E_3^e$ are given by the following expressions [40]:

$$\Pi_3(E_3^e) = -\left(\frac{\partial P_3}{\partial T}\right)_{E_3^e}, \tag{8}$$

$$\Delta T_{EC}(E_3^e) \cong T \int_0^{E_3^e} \frac{1}{\rho_P C_P} \Pi_3 \, dE \approx \frac{T}{\rho C_P} \left( \frac{\alpha_T}{2} [P_3^2(E_3^e) - P_3^2(0)] + \frac{\beta_T}{4} [P_3^4(E_3^e) - P_3^4(0)] + \right.$$
$$\left. \frac{\gamma_T}{6} [P_3^6(E_3^e) - P_3^6(0)] \right), \quad (9)$$

where $\alpha_T = \frac{\partial \alpha_R}{\partial T}$, $\beta_T = \frac{\partial \beta_R}{\partial T}$, and $\gamma_T = \frac{\partial \gamma}{\partial T}$. For the case when $E_3^e$ is equal to the coercive field $E_c$, Eq.(9) contains several contributions to the EC effect, which are proportional to the even powers of the spontaneous polarization $P_s$ and the factor $\frac{T}{\rho C_P}$:

$$\Delta T_{EC}(E_c) \approx -\frac{T}{\rho C_P} \left( \frac{\alpha_T}{2} P_s^2 + \frac{\beta_T}{4} P_s^4 + \frac{\gamma_T}{6} P_s^6 \right); \quad (10a)$$

the heat difference is:

$$\Delta Q_{EC}(E_c) \approx -T \left( \frac{\alpha_T}{2} P_s^2 + \frac{\beta_T}{4} P_s^4 + \frac{\gamma_T}{6} P_s^6 \right). \quad (10b)$$

Let us underline that Eqs.(10) are valid only for a single-domain quasi-homogeneous distribution of $P_3$, because the derivation of the right-hand side in Eq.(9), given in Ref.[40], accounts for neither the domain structure appearance nor the possible polarization rotation in the core-shell BaTiO$_3$ nanorods. Also, it is necessary to consider the heat dissipation and temperature gradient in the case of significant $\vec{r}$-dependence of $P_3(\vec{r}, E_3^e, T)$ for realistic thermal boundary conditions, as well as consider that all experimental measurements are performed at the finite rate of the temperature change (e.g., in adiabatic conditions). To describe the real experimental measurements of the EC effect, it is necessary to solve the thermal problem taking into account the finite rate of the heat transfer and the non-uniform temperature distribution in a multilayer and/or multidomain system (see, e.g., Ref.[41]).

### C. Quasi-static polarization, tetragonality, and electrocaloric properties of single-domain ferroelectric nanowires

Analytical results, calculated using Eqs.(1)-(10) and presented in the subsection, are visualized in Mathematica 13.2 [42]. The calculations were performed for single-domain BaTiO$_3$ core-shell nanowires ($R_c/L_c \ll 0.01$) with different relative shell volumes ($0 \leq \delta V \leq 1$) over a wide temperature range (0 – 1000) K. Due to the virtual absence of the depolarization field in the very long or infinite single-domain nanowire, the effective screening length value does not influence the polar properties, tetragonality, and EC response shown in **Figs. 2 - 4**.

The dependence of the core spontaneous polarization $P_s$ on the relative shell volume $\delta V$ and chemical strain $\delta u$ calculated at room temperature is shown in **Fig. 2(a)**. It is seen that the compressive strains ($\delta u < 0$) suppress the spontaneous polarization, and tensile strains ($\delta u > 0$) induce and enhance the spontaneous polarization. The increase of $\delta V$ leads to the $P_s$ increase for $\delta u > 0$, and supports the paraelectric state for $\delta u \leq 0$. The temperature dependence of $P_s$ calculated for different values of $\delta V$, negative, zero, and positive $\delta u$ are shown in **Figs. 2(b)- 2(f)**. Purple curves, which correspond to $\delta V = 0$ (no shell), coincide with the temperature dependence of the spontaneous polarization of an unstrained bulk BaTiO3. The bulk polarization monotonically decreases from 30 μC/cm² for $T = 0$ to 0 for $T > T_{FE}$, which corresponds to the paraelectric phase. The bulk ferroelectric-paraelectric transition temperature, $T_{FE} \approx 405$ K, is ≈20 K greater than $T_c \approx 383$ K. Red curves, which correspond to $\delta V = 1$ (no core), show the maximal strain-induced changes of $P_s$. It is seen from **Figs. 2(b)-2(d)** that $P_s$ monotonically decreases with $\delta V$ increasing for $\delta u \leq 0$. The monotonic trend gradually disappears with an increase of $\delta u$, and the curves order changes for some "critical" value, $\delta u = \delta u_{cr}$, which can be estimated from the condition $\alpha_R = 0$ for $T = T_c$. From Eq.(7a) for $\alpha_R$, the value $\delta u_{cr}$ depends on the nanorod aspect ratio $R_c/L_c$, length $L_c$, and effective screening length $\lambda$, and is approximately equal to 0.25 %. Indeed, it is seen from **Fig. 2(e)**, where $\delta u = 0.3\%$, that $P_s$ decreases with a $\delta V$ increase for $T < 400$ K, and then increases with a $\delta V$ increase for $T > 400$ K. For $\delta u \gg \delta u_{cr}$, $P_s$ monotonically increases with a $\delta V$ increase (see e.g., **Fig. 2(f)** for $\delta u = 1.5\%$), and the ferroelectric-paraelectric transition temperature increases up to 900 K with a $\delta V$ increase.

The dependence of the core tetragonality ratio $\frac{c}{a}$ on the relative shell volume $\delta V$ and chemical strain $\delta u$ calculated at room temperature is shown in **Fig. 3(a)**. It is seen that compressive strains decrease the tetragonality ratio; and the tetragonality increases for tensile strains. The increase of $\delta V$ leads to the increase of $\frac{c}{a}$ ratio for $\delta u > 0$, and the ratio decreases with an increase of $\delta V$ for $\delta u \leq 0$. The temperature dependence of $\frac{c}{a}$ calculated for different values of $\delta V$, negative, zero, and positive chemical strains $\delta u$ are shown in **Figs. 3(b)-3(f)**. Purple curves, which correspond to $\delta V = 0$ (no shell), coincide with the temperature dependence of $\frac{c}{a}$ ratio for an unstrained bulk BaTiO3, which monotonically decreases from 1.014 for $T = 0$ to 1 for $T \rightarrow T_{FE}$. Red curves, which correspond to $\delta V = 1$ (no core), show the maximal strain-induced change of $\frac{c}{a}$. It is seen from **Figs. 3(b)-3(d)** that $\frac{c}{a}$ monotonically decreases with an increase of $\delta V$ for $\delta u \leq 0$. It is seen from **Fig. 3(e)**, where $\delta u = 0.3\%$, that $\frac{c}{a}$ decreases with a $\delta V$ increase for $T < 400$ K, and then increases with a $\delta V$ increase for $T > 400$ K. For $\delta u \gg \delta u_{cr}$, the core tetragonality monotonically increases with a $\delta V$ increase, becoming weakly temperature-dependent and reaches the maximal value 1.015 for $\delta V = 1$ [see e.g., **Fig. 3(f)** for $\delta u = 1.5\%$].

The dependence of the core EC temperature change $\Delta T_{EC}$ on the relative shell volume $\delta V$ and chemical strain $\delta u$ calculated at room temperature is shown in **Fig. 4(a)**. It is seen that compressive strains decrease the EC cooling effect, which corresponds to $\Delta T_{EC} < 0$; and tensile strains significantly increase the magnitude of $\Delta T_{EC} < 0$. The increase of $\delta V$ leads to the increase of negative $\Delta T_{EC}$ for $\delta u > 0$, and supports the paraelectric state with $\Delta T_{EC} = 0$ for $\delta u \leq 0$. The temperature dependence of $\Delta T_{EC}$ calculated for different values of $\delta V$, negative, zero, and positive chemical strains $\delta u$ are shown in **Fig. 4(b)-4(f)**. Purple curves, which correspond to $\delta V = 0$, are the $\Delta T_{EC}$ values of an unstrained bulk BaTiO$_3$, which reach the maximal value –(3.4 – 3.8) K in the temperature range (280 - 380) K. Red curves, which correspond to $\delta V = 1$ (no core), show the maximal strain-induced changes of $\Delta T_{EC}$, which can reach -5.8 K for the tensile strain +1.5%. It is seen from **Figs. 4(b)-4(d)** that $|\Delta T_{EC}|$ maximum monotonically decreases with a $\delta V$ increase for $\delta u \leq 0$. It is seen from **Fig. 4(e)**, where $\delta u = 0.3\%$, that the EC cooling decreases with a $\delta V$ increase for $T < 400$ K, and then increases with a $\delta V$ increase for $T > 400$ K. For $\delta u \gg \delta u_{cr}$, the EC cooling effect monotonically increases with a $\delta V$ increase and exists up to 900 K, which is the ferroelectric-paraelectric transition temperature for $\delta V \to 1$ [see e.g., **Fig. 4(f)** for $\delta u = 1.5\%$].

Analytical results, shown in **Figs. 2-4**, demonstrate that the synergy of electrostriction coupling and tensile chemical strains can significantly increase the ferroelectric-paraelectric temperature (up to 900 K in comparison with 400 K for bulk BaTiO$_3$), tetragonality (up to +1.015 in comparison with 1.011 for bulk BaTiO$_3$), and EC cooling (up to -6 K in comparison with -3 K for bulk BaTiO$_3$) of the core-shell BaTiO$_3$ nanowires with co-axial polarization. We would like to underline that the results should also be valid for the very long single-domain nanorods whose ends are well-screened. However, the single-domain state should become unstable, as well as the polarization vector should rotate in the unscreened and/or not very long core-shell nanorods. Thus, the analytical results, derived in this section, require numerical verification (especially in the case $\lambda \geq 0.05$ nm and $R_c/L_c > 0.1$). Corresponding FEM results are presented in the next section.

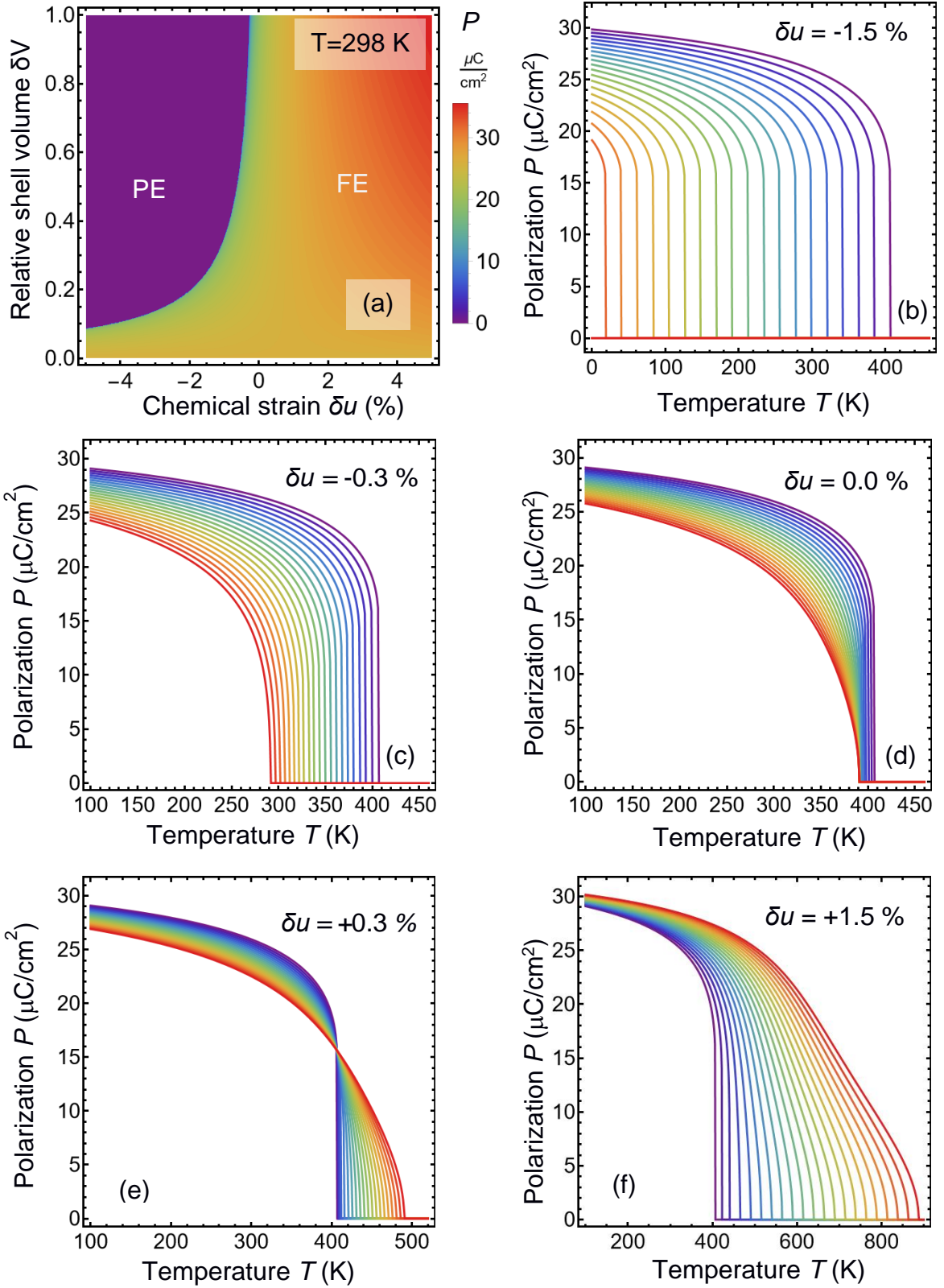

**FIGURE 2.** (a) The dependence of the BaTiO$_3$ nanowire spontaneous polarization on the relative shell volume $\delta V$ and chemical strain $\delta u$ calculated at room temperature $T$ =298 K, $R_c = 10$ nm, and $R_c/L_c \leq 10^{-3}$. Color scale is the polarization in μC/cm$^2$. (b-e) The spontaneous polarization dependence on temperature $T$ calculated for different values of $\delta V$ varying from 0 (purple curves) to 1 (red curves) with a step of 0.05; and chemical strains $\delta u$ =-1.5% (b), -0.3 % (c), 0 (d), 0.3 % (e), and 1.5% (f).

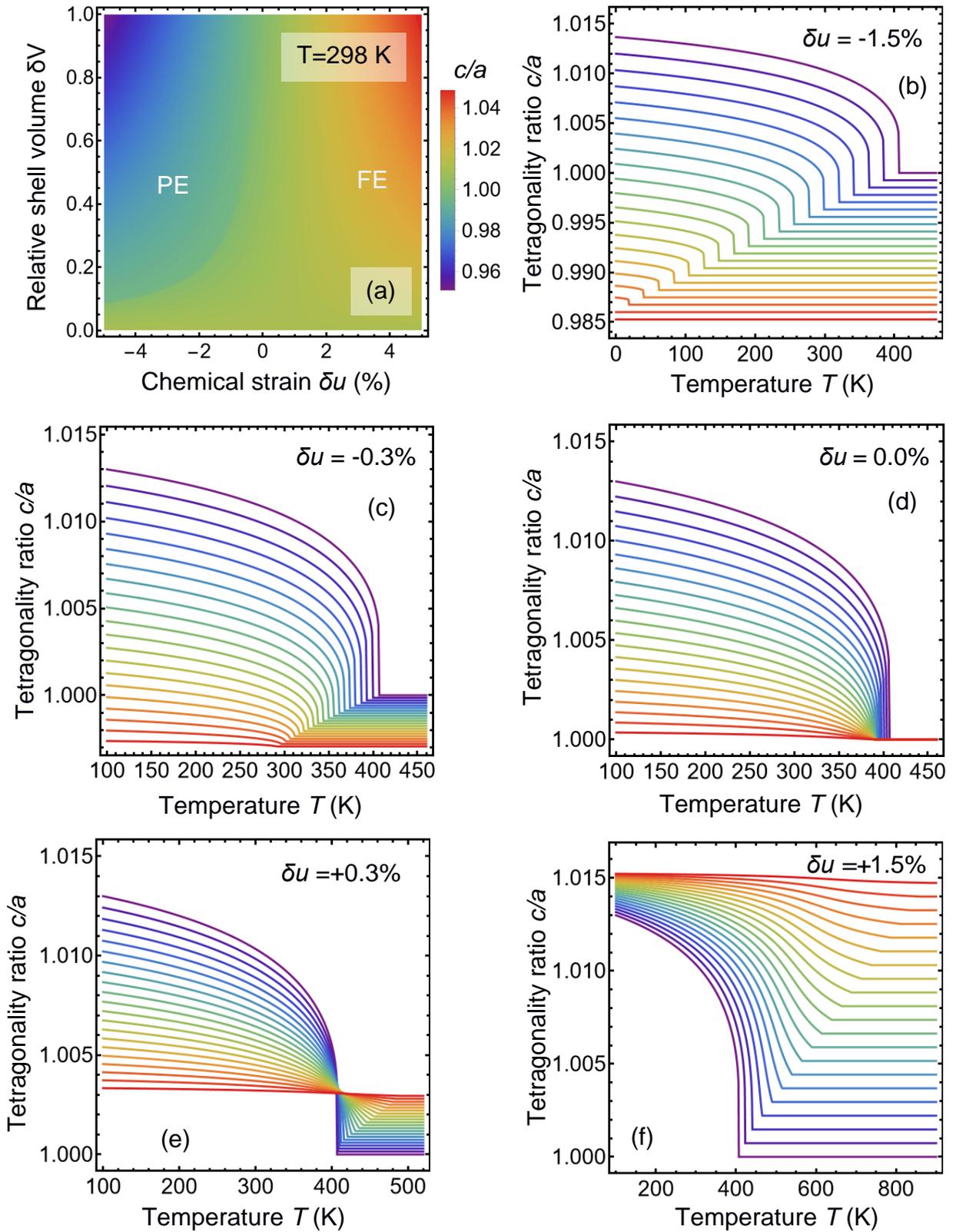

**FIGURE 3. (a)** The dependence of the BaTiO$_3$ nanowire tetragonality ratio $c/a$ on the relative shell volume $\delta V$ and chemical strain $\delta u$. Color scale is the ratio $c/a$. **(b-e)** The tetragonality dependence on temperature $T$ calculated for different values of $\delta V$ varying from 0 (purple curves) to 1 (red curves) with a step of 0.05; and chemical strains $\delta u$ =-1.5% **(b)**, -0.3 % **(c)**, 0 **(d),** 0.3 % **(e),** and 1.5% **(f)**. Other parameters are the same as in **Fig. 2.**

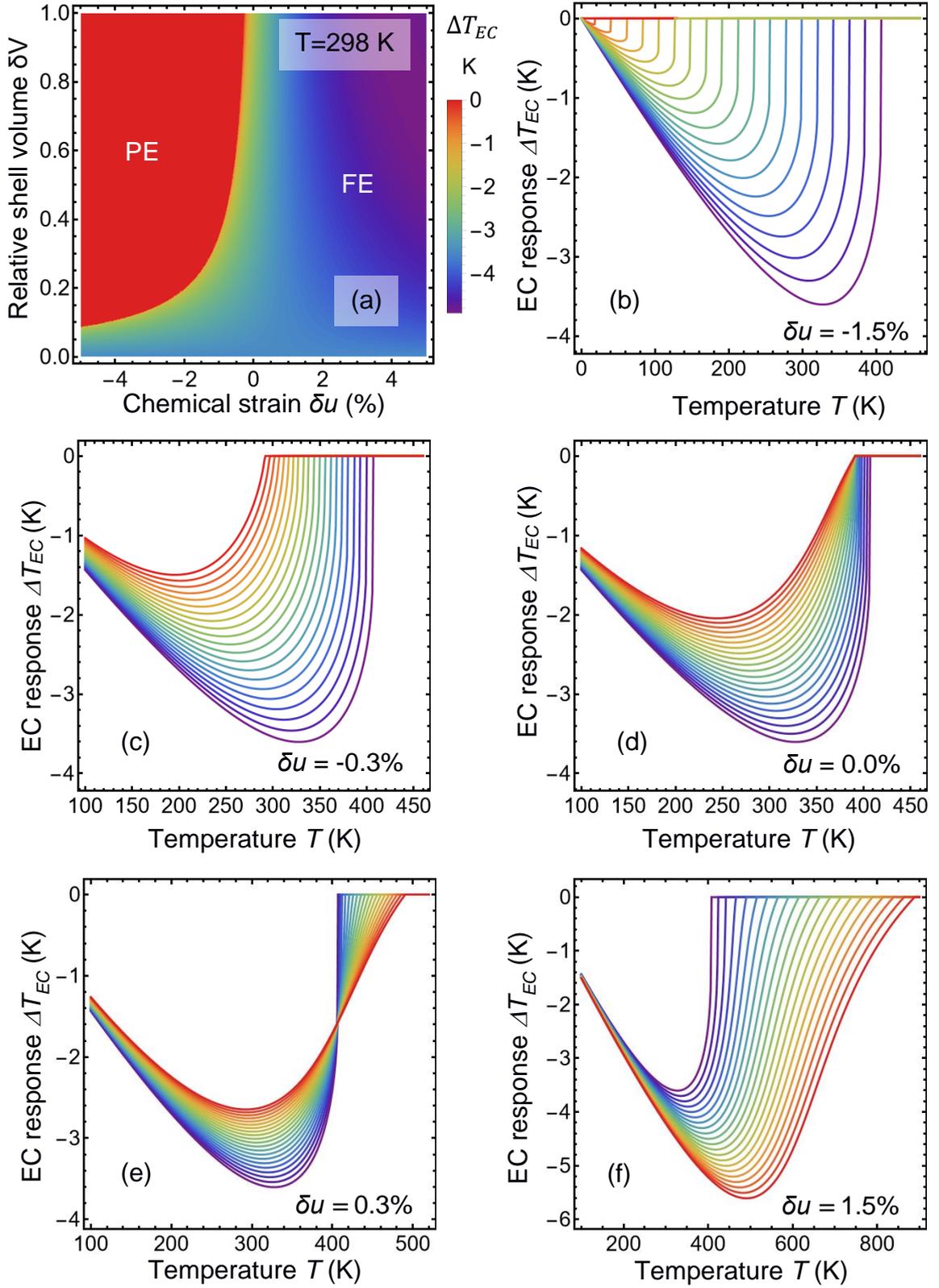

**FIGURE 4. (a)** The dependence of the BaTiO$_3$ nanowire EC temperature change $\Delta T_{EC}$ on the relative shell volume $\delta V$ and chemical strain $\delta u$. Color scale is the temperature change $\Delta T_{EC}$ in Kelvin. **(b-e)** The dependence of $\Delta T_{EC}$ on temperature $T$ calculated for different values of $\delta V$ varying from 0 (purple curves) to 1 (red curves) with a step of 0.05; and chemical strains $\delta u$ = -1.5% **(b)**, -0.3 % **(c)**, 0 **(d),** 0.3 % **(e),** and 1.5% **(f)**. Other parameters are the same as in **Fig. 2**.

## III. FINITE ELEMENT MODELING

The FEM is performed in COMSOL@MultiPhysics software. The COMSOL@MultiPhysics model uses the electrostatics module for the solution of the Poisson equation, solid mechanics, and general math (PDE toolbox) modules for the self-consistent solution of time-dependent LGD equations listed in **Table AI** in **Appendix A** [31]. FEM is performed for different discretization densities of the self-adaptive tetragonal mesh, and randomly small initial polarization distributions. The size of the computational region is not less than 40×40×160 nm$^3$. Material parameters of BaTiO$_3$ are listed in **Table AI** in **Appendix A** [31]. The minimal size of a tetrahedral element in a mesh with fine discretization is equal to the unit cell size, 0.4 nm, the maximal size is (0.8 – 1.2) nm, and 4 nm in the dielectric medium outside the nanorod. The dependence on the mesh size is verified by increasing the minimal size to 0.8 nm. We verified that this results in minor changes in the electric polarization, electric field, and elastic stress and strain, such that the spatial distribution of each of these quantities becomes less smooth (i.e., they contain numerical errors in the form of a small random noise).

FEM are performed for cylindrical core-shell nanorods of different sizes (5 nm < $R_c$ < 25 nm, 20 nm < $L_c$ < 100 nm) and aspect ratios ($R_c/L_c \geq 0.1$). The corresponding geometry of the nanorod is shown in **Fig. 1.** We postulated that the elastic defects are concentrated in an ultra-thin shell layer of thickness 2 nm $\leq \Delta R \leq$ 5 nm under the surface of the nanorod, and the corresponding chemical strains obeys the linear Vegard law. The magnitude of the chemical strains $\delta u$ varies from -3 % to +3 %. The effective screening length $\lambda$ in the shell varies from 0.01 nm to 1 nm. As a rule, the increase of $\lambda$ above 0.1 nm leads to an instability of the single-domain state and induces the formation of various domain morphologies, most interesting of which are discussed below. Stable structures were obtained after a long simulation time, $t \gg 10^3 \tau$, where the parameter $\tau$ is the Landau-Khalatnikov relaxation time, $\tau = \Gamma/|\alpha(0)|$.

The distributions of spontaneous polarization, strain components, and tetragonality $c/a$ in defect-free, tensiled and compressed (by elastic defects) BaTiO$_3$ nanorods are shown in **Fig. 5**. Chemical strains are absent for the top row (a), where $\delta u = 0$, localized under the side surface of the rod in the 2 nm thick shell layer, being equal to $\delta u = +1\%$ for the middle row (b), and $\delta u = -1\%$ for the bottom row (c).

The axial polarization component $P_3$ tends to align along the rod axis in the central part of the defect-free nanorod, and the lateral components, $P_1$ and $P_2$, are almost absent in the central region [see **Fig. 5(a)** for $\delta u = 0$]. The axial strain component $u_3$ is maximal (~1%) in the middle of the defect-free nanorod, and the lateral strain components, $u_1$ and $u_2$, are much smaller (~0.2%) in the region. The strain and polarization behaviors determine the tetragonality (see e.g., Eq.(5)), and therefore $c/a$ is maximal in the central part of the nanorod reaching the value 1.01 in the region. The lateral polarization components, $P_1$ and $P_2$, form an inversely polarized 90-degree vertex-type domain structure near the top and bottom ends of the nanorod. The strains $u_1$ and $u_2$ have a localized maximum, $u_3$ and tetragonality have a localized minimum near the ends.

The axial polarization component $P_3$ tends to align along the rod axis in the central part of the nanorod tensiled by elastic defects in the shell, and the lateral components, $P_1$ and $P_2$, are almost absent in the central region [see **Fig. 5(b)** for $\delta u = +1\%$]. The strain component $u_3$ is maximal in the middle of the rod, and strain components, $u_1$ and $u_2$, are almost absent in the region. The strains $u_1$ and $u_2$ are maximal in the tensiled shell. The tetragonality $c/a$ is maximal in the central part of the rod, where it varies in the range (1.011 - 1.015). The lateral polarization components, $P_1$ and $P_2$, form a distorted meron-like domain structure near the top and bottom ends of the nanorod. The strains $u_1$ and $u_2$ have a diffuse maximum near the ends. The strain component $u_3$ and tetragonality have a diffuse minimum near the ends, except for the shell region where they reach maximal values, 2% and 1.015, respectively.

The spontaneous polarization tends to align perpendicular to the rod axis in the central part of the nanorod compressed by elastic defects in the shell [see **Fig. 5(c)** for $\delta u = -1\%$]. The strain component $u_3$ is small (~ -0.4%) in the middle of the rod, and strain components, $u_1$ and $u_2$, can be significantly higher (up to +0.8%) near the rod axis. Because of this, the tetragonality is minimal (~0.990) in the central part of the rod and near the ends and reaches the highest values (1.001 - 1.005) in the compressed shell. The lateral polarization components, $P_1$ and $P_2$, form a classical vortex-type domain structure entire the nanorod. The strains $u_1$ and $u_2$ have a localized maximum, $u_3$ and tetragonality have a localized minimal near the ends of the nanorod.

The characteristic features of polarization vector morphology in the middle and near the ends of the core-shell nanorod are shown in **Fig. 6** in the form of arrow fields in the lateral $\{x_1, x_2\}$ cross-sections. **Figure 7** shows corresponding distributions of the radial polarization component, $P_r$. It is seen from **Fig. 6**, that the polarization vector becomes curled and forms the vertex-like or chiral meron-like structures near the rod ends, or vortex-like structure in the rod volume, in dependence on the chemical strain magnitude in the shell. Analytical calculations and FEM results, performed in Ref. [43] for strain-free unscreened BaTiO$_3$ nanorods (i.e., for $\lambda \to \infty$ and $\delta u = 0$), reveal the similar chiral meron-like structures near the rod ends, which axial polarization has the flexoelectric nature. They termed them "flexon" because a change of the flexoelectric coefficient sign leads to a reorientation of their axial polarization. FEM performed in this work for tensiled screened BaTiO$_3$ nanorods (i.e., for 0.01 nm $\leq \lambda \leq$ 1 nm and 0.3% $\leq \delta u \leq$ 3%) proved that the flexoelectric coupling determines the meron-like structures chirality and related domain morphology.

The curled structures in the system tend to minimize the free energy consisting of the negative Landau energy, and from the positive polarization gradient energy, elastic and depolarization field energies (see **Appendix A** [31] for details and Refs.[43, 44]). The negative Landau energy is maximal and the positive polarization gradient energy is minimal in the single-domain state of the nanorod. The elastic energy can significantly increase or decrease (which is dependent on the sign and value of the chemical strain $\delta u$) the Landau energy due to the electrostriction coupling. For instance, see Eqs.(7) for qualitative understanding of the Landau energy coefficients renormalization by the strains. The domain formation, which leads to the decrease of the depolarization field divergency, simultaneously decreases the positive depolarization field energy. The polarization screening is incomplete near the rod ends (even for relatively small $\lambda = 0.1$ nm), and the depolarization field is maximal in the spatial regions. The curled domain structures, which emerge near the ends of the rod for all considered $\delta u$, minimize the positive energy of the depolarization electric field. Since the length of the rod is 3 times bigger than its width, the vortices vanish approaching the central part of defect-free and tensiled rods, where the negative Landau energy dominates for $\delta u > 0$. At the same time, the vortices fill the core of the compressed rods, where the negative Landau energy is much smaller for $\delta u < 0$.

Hence, the FEM reveals that the chemical strain in the shell can induce vertex-like (see **Fig. 6(a)** and **7(a)**), meron-like (see **Fig. 6(b)** and **7(b)**), or vortex-like (see **Fig. 6(c)** and **7(c)**) transitions of domain structure morphology in the nanorod core. In particular, tensile chemical strains induce and support the single-domain state in the central part of the nanorod core, meanwhile the curled domain structures appear near the unscreened or partially screened ends of the rod. The vortex-like domains propagate towards the central part of the rod and fill it entirely, when the rod is covered with the compressed shell.

FEM results shows that the vortex intergrowth occurs for chemical compressive strains above some critical value, $\delta u_{VR}^{cr}$, which depends on the temperature, nanorod sizes, aspect ratio and screening conditions at the nanorod ends. The value $\delta u_{VR}^{cr}$ can be estimated as:

$$\delta u_{VR}^{cr} \approx \frac{1}{\delta V}\left(\alpha_T(T-T_c) + \frac{n_d}{2\varepsilon_0[\varepsilon_b n_d + \varepsilon_s(1-n_d) + \varepsilon_s n_d(L_c/\lambda)]}\right)\frac{s_{11}+s_{12}}{Q_{11}+Q_{12}}. \qquad (11)$$

From Eq.(11), larger $\delta V$ (i.e., thicker shells) decreases $\delta u_{VR}^{cr}$. Since we consider the case $T < T_c$, the first term in Eq.(11) is negative and the second term is always positive. Thus, the condition $\delta u_{VR}^{cr} = 0$ becomes valid for a definite nanorod aspect ratio and length (assuming the fixed effective screening length and temperature). Hence, the shape and strain changes allow the control of the domain morphology in the core-shell nanorods of multiaxial ferroelectrics.

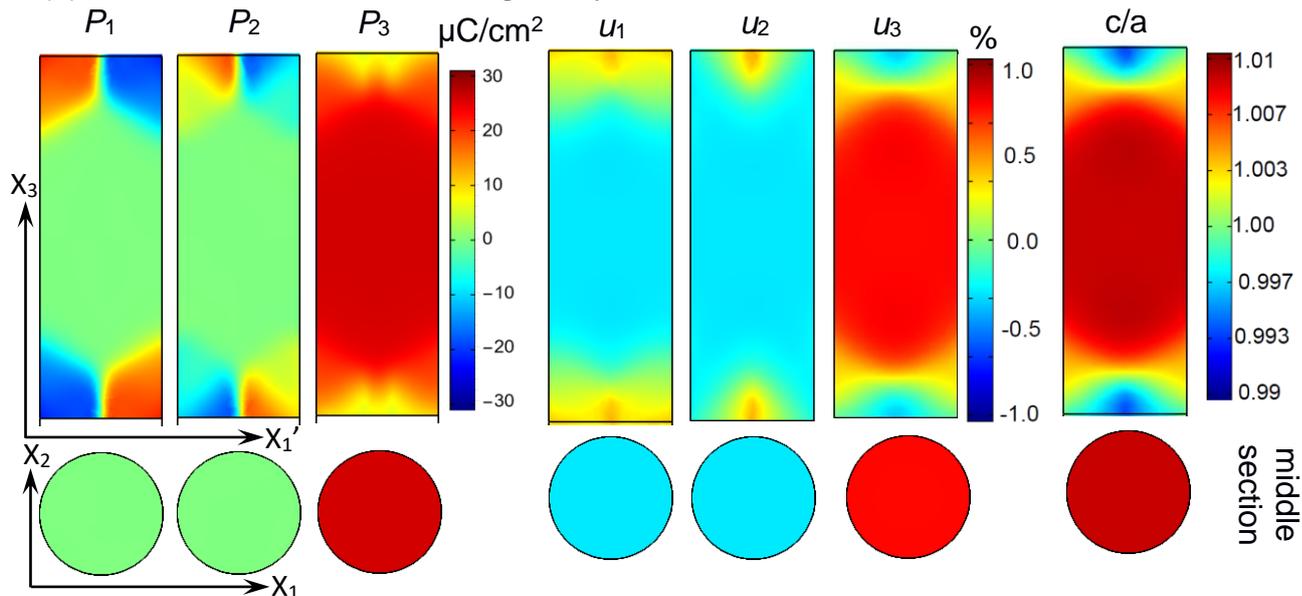
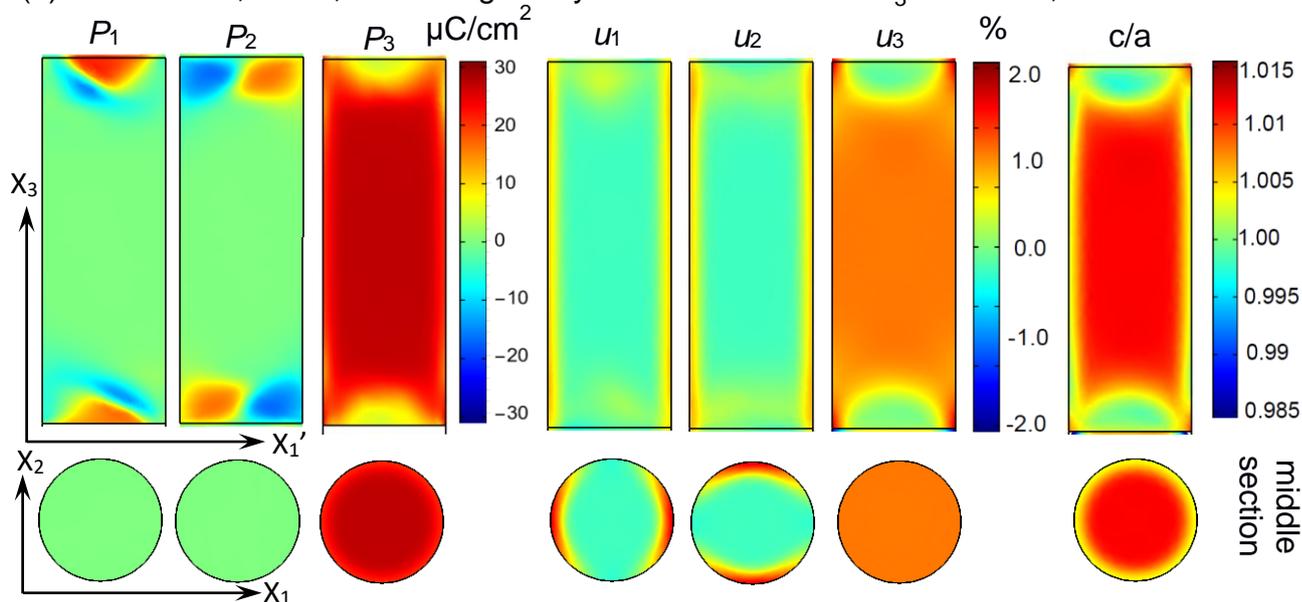
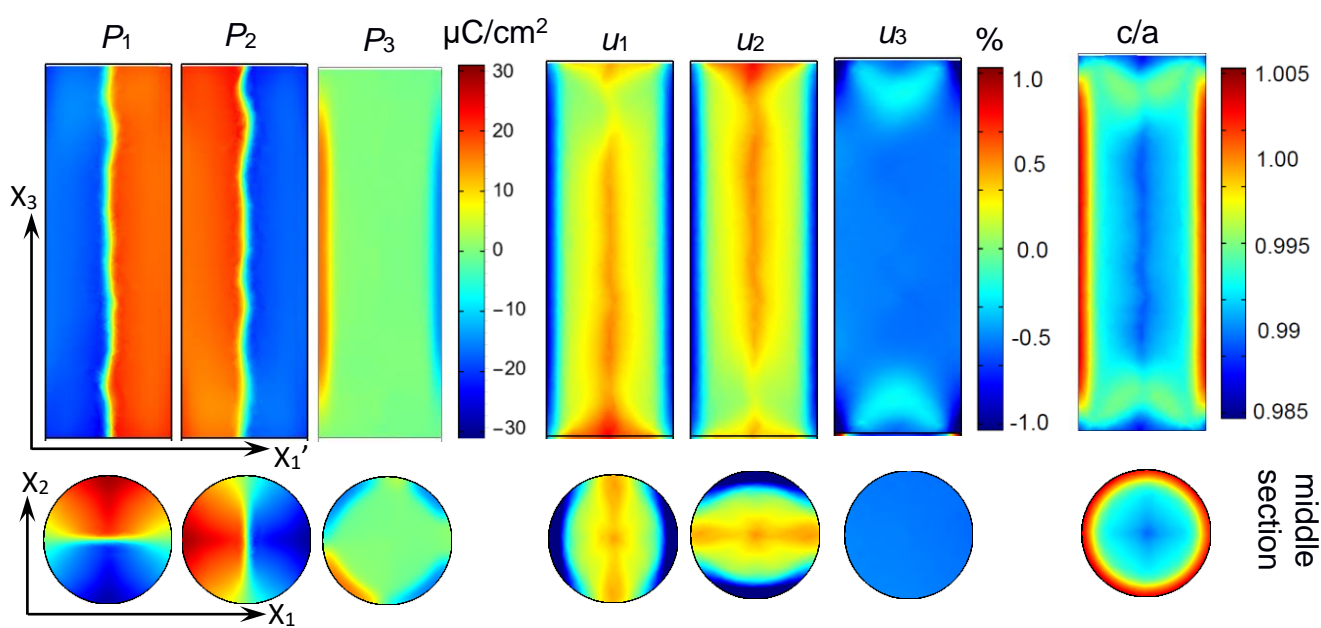

**FIGURE 5.** The distribution of spontaneous polarization components, strains, and tetragonality c/a in the defect-free **(a)**, tensiled **(b)**, and compressed **(c)** core-shell BaTiO$_3$ nanorods**.** Color scales are the polarization components in μC/cm$^2$, strain components in %, and tetragonality in dimensionless units. Chemical strains are absent for the top row **(a)**, where $\delta u = 0$. Chemical strains are localized under the side surface of the rod in the 2 nm thick shell, being equal to $\delta u = +1\%$ for the middle row **(b)**, and $\delta u = -1\%$ for the bottom row **(c)**. The rod radius is 10 nm, the length is 60 nm, the screening length is 0.1 nm, and the temperature $T =$298 K. The "rotated" coordinate $x_1' = (x_1 - x_2)/\sqrt{2}$.

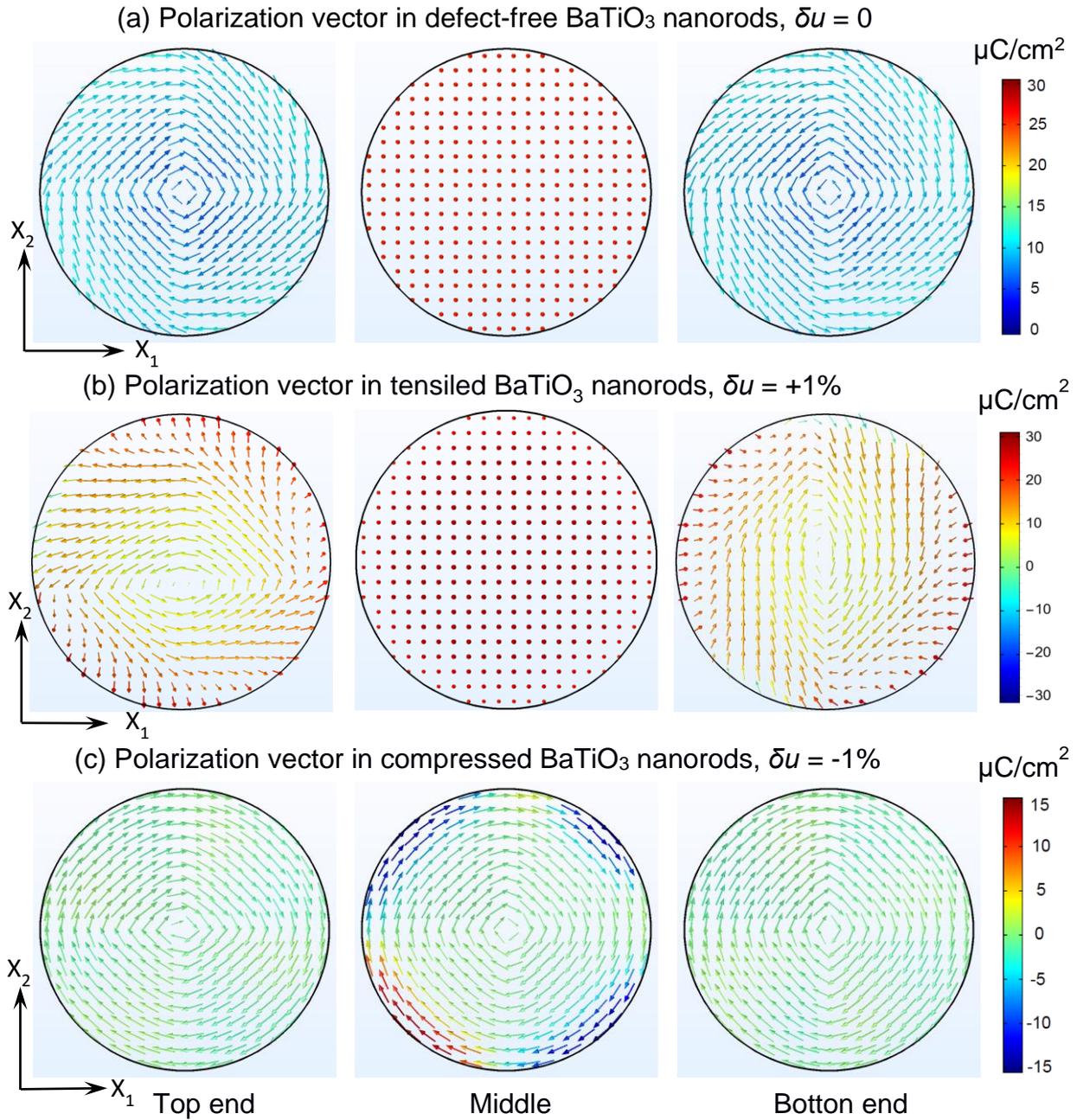

**FIGURE 6.** The distribution of spontaneous polarization vector in the defect-free **(a)**, tensiled **(b)**, and compressed **(c)** core-shell BaTiO$_3$ nanorods**.** Arrows show the orientation of the polarization vector, and their color scale shows the polarization component $P_3$ in μC/cm$^2$. Chemical strains are absent for the top row **(a)**,

where $\delta u = 0$, localized under the side surface of the rod in the 2 nm thick shell layer, being equal to $\delta u = +1\%$ for the middle row **(b)** and $\delta u = -1\%$ for the bottom row **(c)**. Other parameters are the same as in **Fig. 5**.

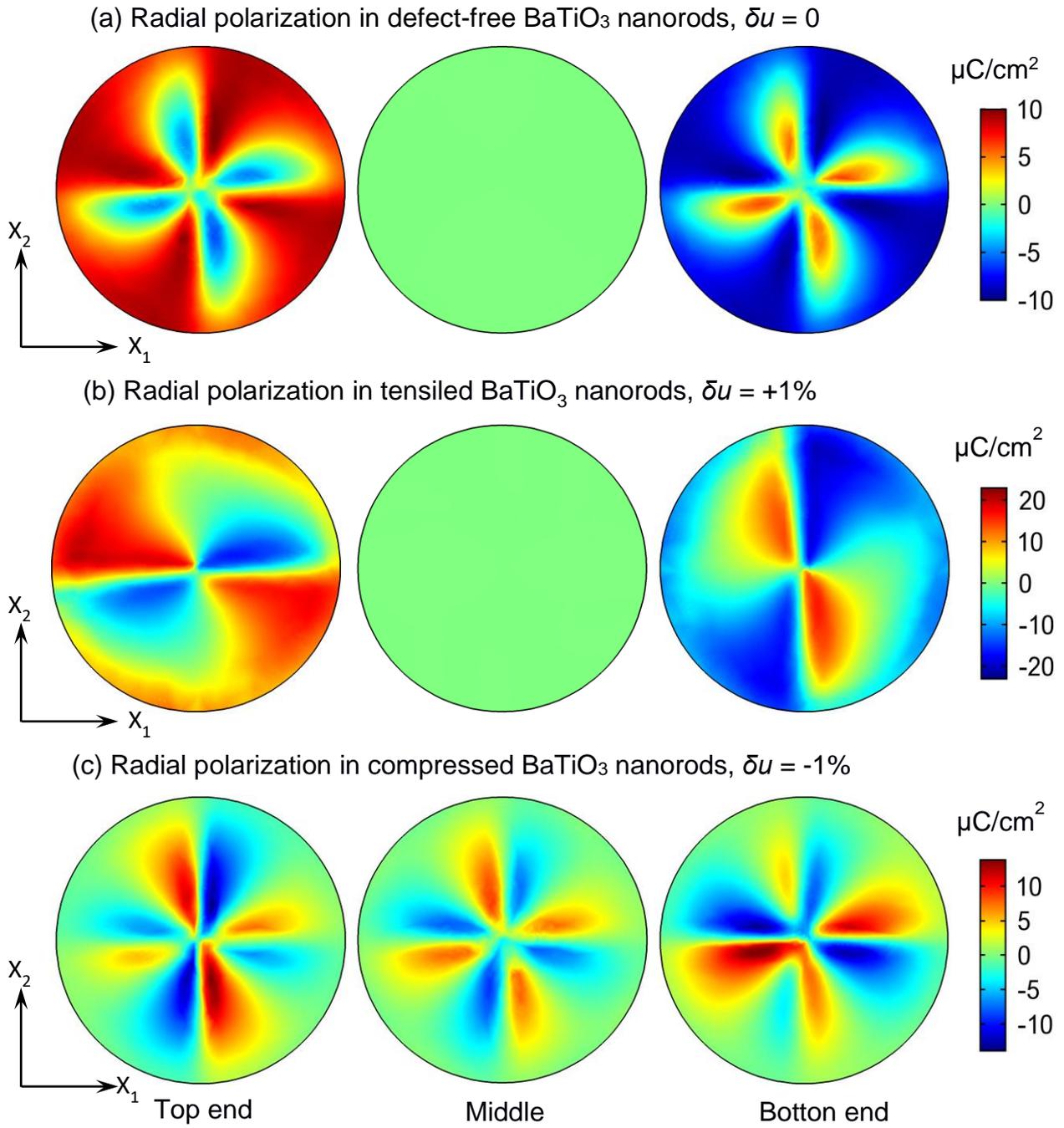

**FIGURE 7.** The distribution of spontaneous polarization radial component $P_r$ in the defect-free **(a)**, tensiled **(b)**, and compressed **(c)** core-shell BaTiO$_3$ nanorods. Color scale shows $P_r$ in µC/cm². Chemical strains are absent for the top row **(a)**, where $\delta u = 0$, localized under the side surface of the rod in the 2 nm thick shell layer, being equal to $\delta u = +1\%$ for the middle row **(b)** and $\delta u = -1\%$ for the bottom row **(c)**. Other parameters are the same as in **Fig. 5**.

To resume, both analytical LGD-based theory and FEM predict that the chemical strains in the shell can increase the nanorod core polarization, lattice tetragonality, and EC cooling effect well-above the values corresponding to the bulk material, as well as the strain control of the domain morphology is possible. The physical reason of the effects is the strong electrostriction coupling between the mismatch-type elastic strains induced in the core by the chemical strains in the shell.

## IV. DISCUSSION
### A. Evidence of tetragonality increase obtained from XRD results

The BaTiO$_3$ nanorods were obtained using a single-step hydrothermal technique and studied by Kovalenko et al. [16]. In the work [16], the phase and structure of the as-prepared BaTiO$_3$ nanopowder were determined using an X-ray diffractometer (XRD) with Cu–Kα radiation. The crystallite size was evaluated based on the size of the coherent scattering region calculated using the Scherrer equation from the full width at half-maximum of the (100) and (001) diffraction peaks. The tetragonality (c/a) was determined by the splitting of (200) peak into (200) and (002) reflections, which are characteristic of the tetragonal structure of BaTiO$_3$. The broadening of the low-angle diffraction lines was used to estimate the sizes of coherent scattering regions, strains, and anisotropy using Williamson-Hall technique. The sizes of the BaTiO$_3$ nanorods were analyzed using a field-emission scanning electron microscope employing a voltage of 3 kV, and the size distribution was obtained from the SEM images.

In this work we refined the XRD data found in Ref. [16], and the lattice constants ratio, $c/a$, appears as high as 1.013 for two powder samples consisting with nanorods, marked as NR1 and NR2, respectively (compare **Table I** in this work with Table I in Ref.[16]). The nanorods average aspect ratio, $R_c/L_c$, is 0.17 and 0.12, their average diameter is 70 nm and 90 nm, and their average length is 410 nm and 770 nm for the samples NR1 and NR2, respectively.

Table I. Characteristic of the BaTiO$_3$ nanorods taken from Ref. [16].

| Sample | Average radius, nm | Average aspect ratio | e$_{001}$ | e$_{100}$ | c/a |
|---|---|---|---|---|---|
| NR1 | 35 | 0.17 | 1.83 | 0.73 | 1.013 |
| NR2 | 45 | 0.12 | 1.67 | 0.66 | 1.013 |

The $c/a$ ratio of the samples is higher than the known value for the bulk BaTiO$_3$ single crystal, $\frac{c}{a} = 1.010$ [45]. Moreover, the high $c/a$ corresponds to different average aspect ratios and radii of the rods, which indicates a weak relation between $c/a$ and the depolarization field effects. The shifting of the diffraction lines (002) and (200) towards lower angles compared to a bulk BaTiO$_3$ was observed for the samples and indicated the lattice expansion due to the presence of OH groups in the crystalline nanorods, apparently into trans-position [46], which leads to the high tetragonality equal to 1.013. The lattice strain in the (001) direction is 2 - 2.5 times greater compared to those in the (100) direction, being unrelated with the nanorod aspect ratio. However, the difference in lattice strains does not affect the degree of anisotropy and tetragonality of the crystalline nanorods, which is consistent with the statement about the effect of OH groups on tetragonality [46]. Furthermore, comparison with the XRD data [16] confirmed the increase of tetragonality ratio in tensiled BaTiO$_3$ nanorods compared to the bulk material.

### B. The negative capacitance effect

It was experimentally demonstrated that in a double-layer capacitor made of paraelectric strontium titanate (SrTiO$_3$) and ferroelectric lead zirconate-titanate (Pb$_x$Zr$_{1-x}$TiO$_3$), the total capacitance is greater than it would be for the SrTiO$_3$ layer of the same thickness as used in the double-layer capacitor [47]. This proves the stabilization of Pb$_x$Zr$_{1-x}$TiO$_3$ in the state of negative differential capacitance (NC) [48]. The NC effect is very important for advanced applications in nanoelectronics [IRDS™ 2021: Beyond CMOS]. Replacing the standard insulator in the gate stack of a field-effect transistor (FET) with a ferroelectric NC insulator of the appropriate thickness has several advantages. The main advantage is that it is a relatively simple replacement for conventional FETs, which significantly reduces heat dissipation of nano-chips with a high density of critical electronic elements.

However, it is very difficult to find the analytical conditions of the NC effect appearance and stability (materials pairs, geometry, temperature and thicknesses ranges) in a general case. Many empirical demonstrations of the NC effect in ferroelectric double-layer capacitors are available [49, 50, 51, 52], and only several works, which contain semi-analytical expressions for the conditions of the NC effect appearance and consider the inevitable appearance of the domain structure in the ferroelectric layer, exist (see e.g., Refs. [53, 54, 55]).

Our analytical calculations and FEM show that ferroelectric BaTiO$_3$ nanorods, which ends are covered by the thin layer (thickness $h_s \leq 10$ nm) of paraelectric SrTiO$_3$, can be suitable candidates for the controllable reduction of the SrTiO$_3$ layer capacitance due to the NC effect emerging in the BaTiO$_3$. Short nanocylinders, e.g., nanopellets (or nanodisks), which length $h_c = 2L_c$ is smaller (or much smaller) than their width $2R_c$ (see **Fig. 8(a)**), are preferable for the capacitor structures miniaturization. In this case the SrTiO$_3$ layers act as a cover for the BaTiO$_3$ core. The physical origin of the NC effect is the specific energy-degenerated metastable states of the spontaneous polarization in BaTiO$_3$ nanocylinders, some examples of which are schematically shown in **Fig. 8(b)**. The free energy potential of these states has relatively flat negative wells, which couple to the positive parabolic potential of the SrTiO$_3$ layers (see read and blue curves in **Fig. 8(c)**). In result, the total potential relief of the BaTiO$_3$ becomes significantly flatter that the SrTiO$_3$ potential, and the charge $Q$ stored at the electrodes covering the three-layer SrTiO$_3$-BaTiO$_3$-SrTiO$_3$ capacitor of the thickness $2h_s + h_c$ can become bigger than the charge $Q_r$ at the electrodes covering the SrTiO$_3$ layer of the thickness $2h_s$. The effective differential capacitance of any electroded system, $C_{eff}$, is equal to the first derivative of the $Q$ over applied voltage $U$, $C_{eff} = \frac{dQ}{dU}$. If the voltage dependence $Q(U)$ is steeper than $Q_r(U)$, the differential capacitance of the SrTiO$_3$-BaTiO$_3$-SrTiO$_3$ capacitor (thickness $2h_s + h_c$) can be greater than the capacitance $C_r = \frac{\varepsilon_0 \varepsilon_s}{2h_s}$ of the reference SrTiO$_3$ capacitor (thickness $2h_s$).

In **Appendix C** [31] we derived that the NC effect exists in the range of thicknesses $h_c$ and $h_s$, chemical strains $\delta u$, shell relative volume $\delta V$, and temperatures $T$, which satisfy the conditions:

$$T - T_c - \frac{\delta u\, \delta V}{\alpha_T} \frac{Q_{11}+Q_{12}}{s_{11}+s_{12}} + \frac{h_s}{\varepsilon_0(\varepsilon_s h_c + 2\varepsilon_b h_s)\alpha_T} > 0, \qquad T - T_c - \frac{\delta u\, \delta V}{\alpha_T} \frac{Q_{11}+Q_{12}}{s_{11}+s_{12}} < 0. \qquad (12)$$

In Eq.(12) we regard that $h_c \ll 2R_c$. The term $\frac{\delta u\, \delta V}{\alpha_T} \frac{Q_{11}+Q_{12}}{s_{11}+s_{12}}$ is the shift of the bulk Curie temperature $T_c$ induced by the chemical strains. The term $\frac{h_s}{\varepsilon_0(\varepsilon_s h_c + 2\varepsilon_b h_s)\alpha_T}$ is the decrease of $T_c$ originated from the depolarization field of a single-domain BaTiO$_3$ core. Hence, the conditions (12) are valid for the BaTiO$_3$ core in the region of size-induced paraelectric (PE) phase coexisting with the "shallow" ferroelectric (FE) phase. Notably that the energy-degenerated metastable domain states occur exactly in the region of PE and FE phases coexistence in the ferroelectrics with the first order FE-PE phase transition. The difference of the three-layer capacitance and reference capacitance is given by the expression:

$$\Delta C = C_{eff} - C_r = \frac{\varepsilon_0 \varepsilon_s}{2h_s} \left( \frac{\frac{h_s}{\varepsilon_s h_c + 2\varepsilon_b h_s}}{\varepsilon_0 \left\{ \alpha_T(T-T_c) - \delta u\, \delta V \frac{Q_{11}+Q_{12}}{s_{11}+s_{12}} \right\} + \frac{h_s}{\varepsilon_s h_c + 2\varepsilon_b h_s}} - 1 \right) \frac{h_c \varepsilon_s}{\varepsilon_s h_c + 2\varepsilon_b h_s}. \qquad (13)$$

The dependence of the dimensionless ratio, $\frac{\Delta C}{C_r}$, on the relative strain $\delta u \delta V$ and thickness ratio $\frac{h_c}{h_s}$ calculated at room temperature is shown in **Fig. 8(d)**. The ratio $\frac{\Delta C}{C_r}$ is negative (which corresponds to $C_{eff} < C_r$) in the lower rectangular region **Fig. 8(d)**. The region corresponds to the PE phase of a bulk BaTiO$_3$. The ratio $\frac{\Delta C}{C_r}$ is zero along the black horizontal line $T = T_c + \frac{\delta u\, \delta V}{\alpha_T} \frac{Q_{11}+Q_{12}}{s_{11}+s_{12}}$ and positive (which corresponds to the NC effect) between the black horizontal line and the black hyperbolae, $\frac{1}{\varepsilon_s(h_c/h_s)+2\varepsilon_b} = \delta u\, \delta V \frac{Q_{11}+Q_{12}}{s_{11}+s_{12}} - \alpha_T(T - T_c)$. The hyperbolae is the boundary between the size-induced PE phase and the single-domain FE phase, and thus $C_{eff}$ sharply increases approaching the PE-FE boundary and diverges ($C_{eff} \to \infty$) at it. Note that the red color in **Fig. 8(d)** corresponds to $\frac{\Delta C}{C_r} \geq 5$, and white color corresponds to the region of the "deep" single-domain FE phase, where $\frac{\Delta C}{C_r} < -1$ and tends to $-\infty$ approaching the FE-PE boundary.

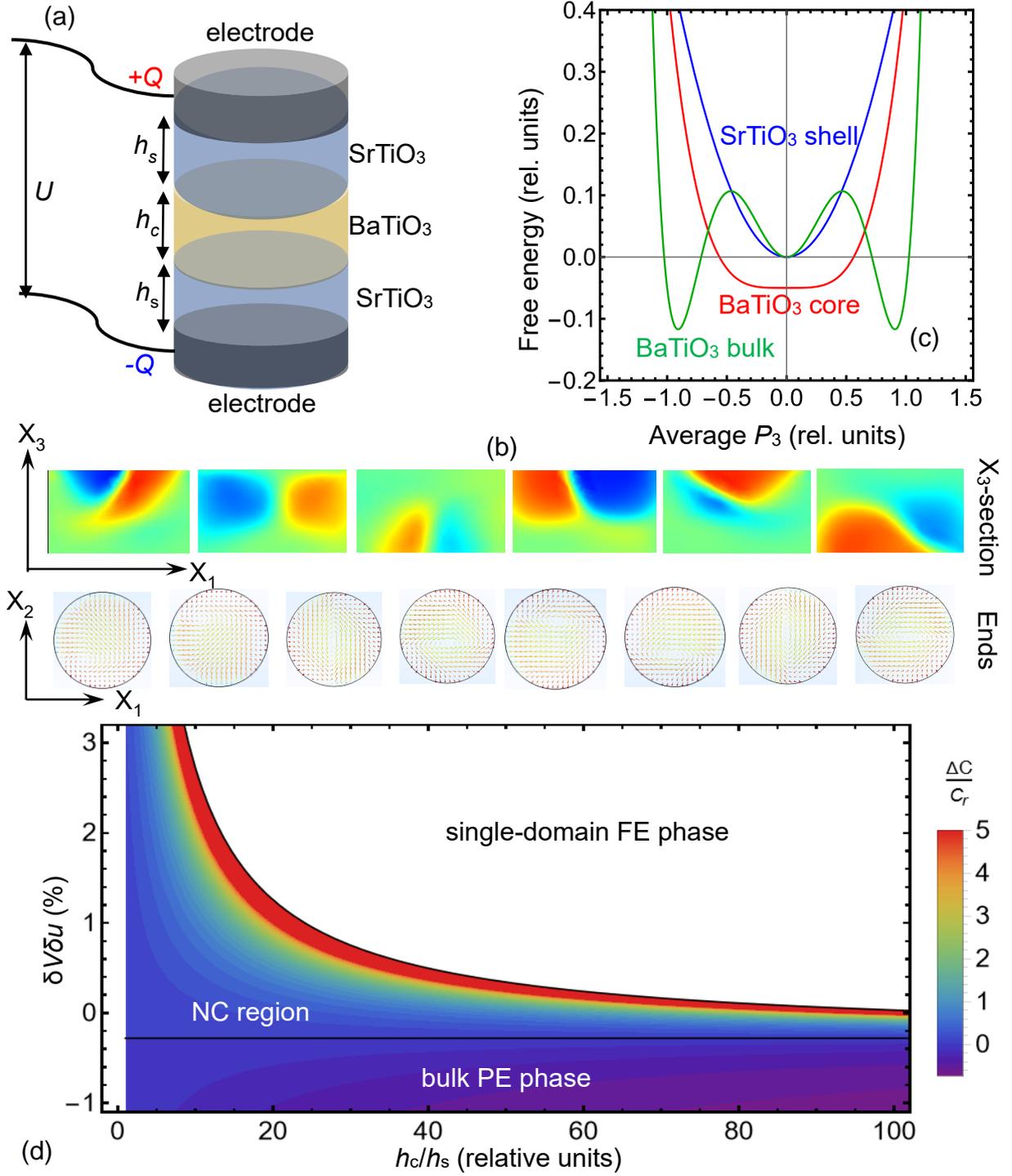

**FIGURE 8.** (a) Three-layer capacitor consisting of the BaTiO$_3$ nanocylinder, which ends are covered by the paraelectric SrTiO$_3$ layers. (b) Typical metastable states of the spontaneous polarization in the SrTiO$_3$-BaTiO$_3$-SrTiO$_3$ nanocapacitor. (c) Schematic illustration of the free energy dependence on the polarization for the single-domain bulk BaTiO$_3$ (the green curve), paraelectric SrTiO$_3$ shell (the blue curve) and the BaTiO$_3$ core with the metastable polarization states (the red curve). (d) The dependence of the dimensionless ratio, $\frac{\Delta C}{C_r}$, on the relative strain $\delta u \delta V$ and thickness ratio $\frac{h_c}{h_s}$ calculated for $T = 298$ K and $\varepsilon_s = 300$, which corresponds to SrTiO$_3$. Color scale is the ratio $\frac{\Delta C}{C_r}$ in dimensionless units. Other parameters are the same as in **Fig. 2**.

It is seen from Eqs.(12)-(13) and **Fig. 8(d)** that the magnitude of the NC effect is controlled by the chemical strain and relative shell volume (namely, by the product $\delta u \delta V$), as well as by the thickness ratio $\frac{h_c}{h_s}$. Since it is relatively easy to change the sizes and geometry of core-shell nanoparticles (i.e., parameters $\delta V$, $h_s$ and $h_c$), they are suitable objects for the NC effect control.

### C. Conclusions

Using the LGD approach, we derive analytical expressions for the spontaneous polarization, tetragonality, and electrocaloric response in core-shell nanorods. We postulate that the nanorod core presents a defect-free single-crystalline ferroelectric material, and the elastic defects are accumulated in an ultra-thin shell, where they can induce tensile or compressive chemical strains.

The FEM reveals the strain-induced transitions of domain structure morphology in the nanorods. Namely, tensile chemical strains induce and support the single-domain state in the central part of the nanorod, while the curled domain structures appear near the unscreened or partially screened ends of the rod. The vortex-like domains propagate towards the central part of the rod and fill it entirely when the rod is covered with the shell compressed by elastic defects. The vortex intergrowth occurs for compressive strains above some critical value, which depends on the nanorod sizes, aspect ratio, and screening conditions at the nanorod ends.

Both analytical theory and FEM predict that the tensile chemical strains in the shell increase of the nanorod polarization, lattice tetragonality, and electrocaloric cooling effect well-above the values corresponding to the bulk material. The physical reason of the increase is the strong electrostriction coupling between the mismatch-type elastic strains induced in the core by the chemical strains in the shell. Comparison with the XRD data published earlier confirmed the increase of tetragonality ratio in tensiled $BaTiO_3$ nanorods compared to the bulk material.

Analytical calculations and FEM show that $BaTiO_3$ nanopellets, which ends are covered by $SrTiO_3$ layers, can be suitable candidates for the controllable the NC effect. Obtained analytical expressions, which are suitable for the description of strain-induced changes in a wide class of multiaxial ferroelectric core-shell nanorods, nanowires and nanopellets, can be useful for prediction and strain engineering of advanced ferroelectric nanomaterials.

**Authors contribution.** A.N.M. generated the research idea, formulated the problem, performed most of analytical calculations and wrote the paper draft. E.A.E. wrote the FEM codes and prepare figures. The treatment of XRD data is performed by O.A.K. D.R.E. worked on the results interpretation, discussion, and paper improvement.

**Acknowledgments.** Authors are very grateful to the Referee for the very useful remarks and suggestions. A.N.M. acknowledges EOARD project 9IOE063b and related STCU partner project P751b. The treatment of XRD data presented in the work (O.A.K.) was supported by the NATO SPS G5980 – FRAPCOM. E.A.E. acknowledges the DOE Software Project on "Computational Mesoscale Science and Open Software for Quantum Materials", under Award Number DE-SC0020145 as part of the Computational Materials Sciences Program of US Department of Energy, Office of Science, Basic Energy Sciences.

# Supplementary Materials to
# "The Influence of Chemical Strains on the Electrocaloric Response, Polarization Morphology, Tetragonality and Negative Capacitance Effect of Ferroelectric Core-Shell Nanorods and Nanowires"

### APPENDIX A. The LGD free energy functional

The LGD free energy functional $G$ of the core polarization $\boldsymbol{P}$ additively includes a Landau expansion on the 2-nd, 4-th, 6-th, and 8-th powers of the polarization, $G_{Landau}$; a polarization gradient energy contribution, $G_{grad}$; an electrostatic contribution, $G_{el}$; the elastic, linear, and nonlinear electrostriction couplings and flexoelectric contributions, $G_{es+flexo}$; and a surface energy, $G_S$. The functional $G$ has the form [1, 2, 3]:

$$G = G_{Landau} + G_{grad} + G_{el} + G_{es+flexo} + G_{VS} + G_S, \quad (A.1)$$

$$G_{Landau} = \int_{0<r<R_c} d^3r \left[ a_i P_i^2 + a_{ij} P_i^2 P_j^2 + a_{ijk} P_i^2 P_j^2 P_k^2 + a_{ijkl} P_i^2 P_j^2 P_k^2 P_l^2 \right], \quad (A.2a)$$

$$G_{grad} = \int_{0<r<R} d^3r \frac{g_{ijkl}}{2} \frac{\partial P_i}{\partial x_j} \frac{\partial P_k}{\partial x_l}, \quad (A.2b)$$

$$G_{el} = -\int_{0<r<R_c} d^3r \left( P_i E_i + \frac{\varepsilon_0 \varepsilon_b}{2} E_i E_i \right) - \frac{\varepsilon_0}{2} \int_{R_c<r<R_s} \varepsilon_{ij}^S E_i E_j d^3r - \frac{\varepsilon_0}{2} \int_{r>R+\Delta R} \varepsilon_{ij}^e E_i E_j d^3r, \quad (A.2c)$$

$$G_{es+flexo} = -\int_{0<r<R_c} d^3r \left( \frac{s_{ijkl}}{2} \sigma_{ij} \sigma_{kl} + Q_{ijkl} \sigma_{ij} P_k P_l + Z_{ijklmn} \sigma_{ij} P_k P_l P_m P_n + \frac{1}{2} W_{ijklmn} \sigma_{ij} \sigma_{kl} P_m P_n + F_{ijkl} \sigma_{ij} \frac{\partial P_l}{\partial x_k} \right), \quad (A.2d)$$

$$G_S = \frac{1}{2} \int_{r=R_c} d^2r \, a_{ij}^{(S)} P_i P_j. \quad (A.2e)$$

The coefficient $a_i$ linearly depends on temperature $T$:

$$a_i(T) = \alpha_T [T - T_C(R_c)], \quad (A.3a)$$

where $\alpha_T$ is the inverse Curie-Weiss constant, and $T_C(R_c)$ is the ferroelectric Curie temperature renormalized by electrostriction and surface tension as [1, 2]:

$$T_C(R_c) = T_C \left( 1 - \frac{Q_c}{\alpha_T T_C} \frac{2\mu}{R_c} \right), \quad (A.3b)$$

where $T_C$ is a Curie temperature of a bulk ferroelectric. $Q_c$ is the sum of the electrostriction tensor diagonal components, which is positive for most ferroelectric perovskites with cubic m3m symmetry in the paraelectric phase, namely $0.005 < Q_c < 0.05$ (in m$^4$/C$^2$). $\mu$ is the surface tension coefficient.

**Table AI.** LGD coefficients and other material parameters of a BaTiO$_3$ core in Voigt notations. Adapted from Ref. [4].

| Parameter, its description, and dimension (in the brackets) | The numerical value or variation range of the LGD parameters |
|---|---|
| Expansion coefficients $a_i$ in the term $a_i P_i^2$ in Eq.(A.2b) (C$^{-2}$·mJ) | $a_1 = 3.33(T-383) \times 10^5$ |
| Expansion coefficients $a_{ij}$ in the term $a_{ij} P_i^2 P_j^2$ in Eq.(A.2b) (C$^{-4}$·m$^5$J) | $a_{11} = 3.6(T-448) \times 10^6$, $a_{12} = 4.9 \times 10^8$ |
| Expansion coefficients $a_{ijk}$ in the term $a_{ijk} P_i^2 P_j^2 P_k^2$ in Eq.(A.2b) (C$^{-6}$·m$^9$J) | $a_{111} = 6.6 \times 10^9$, $a_{112} = 2.9 \times 10^9$, $a_{123} = 3.64 \times 10^{10} + 7.6(T-293) \times 10^{10}$. |
| Expansion coefficients $a_{ijkl}$ in the term $a_{ijkl} P_i^2 P_j^2 P_k^2 P_l^2$ in Eq.(A.2b) (C$^{-8}$·m$^{13}$J) | $a_{1111} = 4.84 \times 10^7$, $a_{1112} = 2.53 \times 10^7$, $a_{1122} = 2.80 \times 10^7$, $a_{123} = 9.35 \times 10^7$. |
| Linear electrostriction tensor $Q_{ijkl}$ in the term $Q_{ijkl}\sigma_{ij}P_k P_l$ in Eq.(A.2e) (C$^{-2}$·m$^4$) | In Voigt notations $Q_{ijkl} \to Q_{ij}$, which are equal to $Q_{11}=0.11$, $Q_{12}=-0.045$, $Q_{44}=0.059$ |
| Nonlinear electrostriction tensor $Z_{ijklmn}$ in the term $Z_{ijklmn}\sigma_{ij}P_k P_l P_m P_n$ in Eq.(A.2e) (C$^{-4}$·m$^8$) | In Voigt notations $Z_{ijklmn} \to Z_{ijk}$. $Z_{ijk}$ varies in the range $-1 \leq Z_c \leq 1$ [5] |
| Nonlinear electrostriction tensor $W_{ijklmn}$ in the term $W_{ijklmn} \sigma_{ij}\sigma_{kl}P_m P_n$ in Eq.(A.2e) (C$^{-2}$·m$^4$ Pa$^{-1}$) | In Voigt notations $W_{ijklmn} \to W_{ijk}$. $W_{ij3}$ varies in the range of $0 \leq W_c \leq 10^{-12}$ as a very small free parameter, and we can neglect it, putting $W_{ij3} = 0$ |
| Elastic compliances tensor, $s_{ijkl}$, in Eq.(A.2e) (Pa$^{-1}$) | In Voigt notations $s_{ijkl} \to s_{ij}$, which are equal to $s_{11}=8.3 \times 10^{-12}$, $s_{12}=-2.7 \times 10^{-12}$, $s_{44}=9.24 \times 10^{-12}$. |
| Polarization gradient coefficients $g_{ijkl}$ in Eq.(A.2c) (C$^{-2}$m$^3$J) | In Voigt notations $g_{ijkl} \to g_{ij}$, which are equal to: $g_{11}=5.0 \times 10^{-10}$, $g_{12}=-0.2 \times 10^{-10}$, $g_{44}= 0.2 \times 10^{-10}$. |
| Flexoelectric coefficients $F_{ijkl}$ in Eq.(A.2d) (10$^{-11}$m$^3$/C) | In Voigt notations $F_{ijkl} \to F_{ij}$, which are equal to $F_{11} = 2.4$, $F_{12} = 0.5$, $F_{44} = 0.06$ (these values are recalculated from Ref.[6]) |
| Surface energy coefficients $a_{ij}^{(S)}$ in Eq.(A.2f) | 0 (that corresponds to the natural boundary conditions) |
| Core radius $R_c$ (nm) | Variable: 5 – 50 |

| Background permittivity $\varepsilon_b$ in Eq.(A.2d) (unity) | 7 |

\* $\alpha = 2a_1, \beta = 4a_{11}, \gamma = 6a_{111}$, and $\delta = 8a_{1111}$

## APPENDIX B. Analytical solution of elastic problem for a ferroelectric nanowire

The free energy expansion on polarization $P_3$ and stress $\sigma_i$ powers has the following form:

$$\Delta F = a_1 P_3^2 + a_{11} P_3^4 + a_{111} P_3^6 - Q_{11}\sigma_3 P_3^2 - Q_{12}(\sigma_1 + \sigma_2)P_3^2 - E_3 P_3 - \frac{1}{2}s_{11}(\sigma_1^2 + \sigma_2^2 + \sigma_3^2) - s_{12}(\sigma_1\sigma_2 + \sigma_1\sigma_3 + \sigma_3\sigma_2) - \frac{1}{2}s_{44}(\sigma_4^2 + \sigma_5^2 + \sigma_6^2) - (\sigma_1 + \sigma_2 + \sigma_3)u_{VT}. \quad (B.1)$$

Here $Q_{\alpha\beta}$ are electrostriction coefficients, $E_3$ is the electric field component along the wire axis, and $u_{VT}$ is the chemical and/or thermal expansion strain. Hereinafter we use the Voigt notations for $\sigma_i$ or matrix notation for $\sigma_{nm}$ ($xx \to 1, yy \to 2, zz \to 3, zy \to 4, zx \to 5$ and $xy \to 6$) when necessary.

Variation of Eq.(B.1) with respect to stress gives "modified Hooke's" law

$$u_1 = s_{11}\sigma_1 + s_{12}(\sigma_2 + \sigma_3) + Q_{12}P_3^2 + F_{12}\frac{\partial P_3}{\partial z} + u_{VT}, \quad (B.2a)$$

$$u_2 = s_{11}\sigma_2 + s_{12}(\sigma_1 + \sigma_3) + Q_{12}P_3^2 + F_{12}\frac{\partial P_3}{\partial z} + u_{VT}, \quad (B.2b)$$

$$u_3 = s_{11}\sigma_3 + s_{12}(\sigma_2 + \sigma_1) + Q_{11}P_3^2 + F_{11}\frac{\partial P_3}{\partial z} + u_{VT}, \quad (B.2c)$$

$$u_4 = s_{44}\sigma_4 + F_{44}\frac{\partial P_3}{\partial y}, \quad (B.2d)$$

$$u_5 = s_{44}\sigma_5 + F_{44}\frac{\partial P_3}{\partial x}, \quad (B.2e)$$

$$u_6 = s_{44}\sigma_6. \quad (B.2f)$$

Similar equations are valid in the diffuse shell except for the absence of the electrostriction term. From general symmetry consideration, we can suggest that the displacement vector has the radial and axial components, $u_\rho(z,\rho)$ and $u_z(z,\rho)$, which depend on the polar radius $\rho$ and axial coordinate z. A general homogeneous solution for the mechanical displacement of a radially symmetric wire is [7]:

$$U_z = u_0 + az, \quad U_\rho = b\rho + \frac{c}{\rho}. \quad (B.3)$$

Here the constants $u_0$, $a$, $b$, and $c$ should be determined from the boundary and/or interfacial conditions. In this case, the components of strain tensor in cylindrical coordinate system are:

$$u_{zz} = \frac{\partial U_z}{\partial z} \equiv a, \quad u_{\rho\rho} = \frac{\partial U_\rho}{\partial \rho} \equiv b - \frac{c}{\rho^2}, \quad u_{\psi\psi} = \frac{U_\rho}{\rho} \equiv b + \frac{c}{\rho^2}, \quad (B.4a)$$

$$u_{z\rho} = \frac{1}{2}\left(\frac{\partial U_\rho}{\partial z} + \frac{\partial U_z}{\partial \rho}\right) \equiv 0, \quad u_{\rho\psi} = 0, \quad u_{z\psi} = 0. \quad (B.4b)$$

It is seen from Eqs.(B.2) that the solution (B.3), obtained for the single-domain rod with a homogeneous polarization, is valid in a general case too, and thus the strain field (B.4) corresponds to the following stress tensor:

$$\sigma_{zz} = A, \quad \sigma_{\rho\rho} = B - \frac{C}{\rho^2}, \quad \sigma_{\psi\psi} = B + \frac{C}{\rho^2}, \quad \sigma_{z\rho} = \sigma_{\rho\psi} = \sigma_{z\psi} = 0. \quad (B.5)$$

Here the constants $A$, $B$, and $C$ should be determined from the boundary and/or interfacial conditions. They are related with the constants $u_0$, $a$, $b$, and $c$ by Eqs. (B.2). Below we apply the general solution (B.4)-(B.5) to the considered physical problem.

### B.1. The core-shell model for long nanorods and nanowires

Let consider a bilayer nanorod which has a ferroelectric core and a paraelectric shell. For the sake of simplicity, we suppose that the core and the shell have the same isotropic elastic compliances tensor. Note that the radially symmetric solution is impossible even for the cubic anisotropy of elastic properties. From Eq.(B.5) the solution for the core can be written as

$$\sigma_{zz}^c = A_c, \quad \sigma_{\rho\rho}^c = B_c, \quad \sigma_{\psi\psi}^c = B_c. \quad (B.6a)$$

Here we omitted the divergent term $\sim 1/\rho^2$ to keep the solution finite. From Eq.(B.5) the solution for the shell can be written as

$$\sigma_{zz}^s = A_s, \quad \sigma_{\rho\rho}^s = B_s - \frac{C_s}{\rho^2}, \quad \sigma_{\psi\psi}^s = B_s + \frac{C_s}{\rho^2}. \quad (B.6b)$$

The strain components for the core and shell of the nanorod are expressed via the stress components as follows:

$$u_{\rho\rho}^{s,c} = s_{11}\sigma_{\rho\rho}^{s,c} + s_{12}(\sigma_{\psi\psi}^{s,c} + \sigma_{zz}^{s,c}) + Q_{12}P_3^2 + u_{s,c}, \quad (B.7a)$$

$$u_{\psi\psi}^{s,c} = s_{11}\sigma_{\psi\psi}^{s,c} + s_{12}(\sigma_{\rho\rho}^{s,c} + \sigma_{zz}^{s,c}) + Q_{12}P_3^2 + u_{s,c}, \quad (B.7b)$$

$$u_{zz}^{s,c} = s_{11}\sigma_{zz}^{s,c} + s_{12}(\sigma_{\psi\psi}^{s,c} + \sigma_{\rho\rho}^{s,c}) + Q_{11}P_3^2 + u_{s,c}. \quad (B.7b)$$

Here only nontrivial components are listed; $u_c$ and $u_s$ are the "effective" (e.g., chemical and/or thermal) strains of the core and shell, respectively. Interfacial and boundary conditions for either continuity (B.8a) or absence of stress (B.8b), have the following form:

$$\sigma_{\rho\rho}^c(\rho = R_c) = \sigma_{\rho\rho}^s(\rho = R_c), \quad (B.8a)$$

$$\sigma_{\rho\rho}^s(\rho = R_s) = -\frac{\mu}{R_s}, \quad \sigma_{zz}(z = \pm h) = 0 \quad (B.8b)$$

Due to the reasons discussed in the Introduction, one can neglect the surface tension contribution ($\sim \frac{\mu}{R_s}$) in Eq.(B.8b). Interfacial conditions of the strain and displacement components continuity at the shell-core interface are:

$$U_\rho^c(\rho = R_c) = U_\rho^s(\rho = R_c) \Leftrightarrow u_{\psi\psi}^c(\rho = R_c) = u_{\psi\psi}^s(\rho = R_c), \quad (B.8c)$$

$$U_z^c(\rho = R_c) = U_z^s(\rho = R_c) \Leftrightarrow u_{zz}^c(\rho = R_c) = u_{zz}^s(\rho = R_c). \quad (B.8d)$$

The condition (B.8b) of stress-free top and bottom ends of the nanorod could not be satisfied with solutions like Eqs. (B.4)-(B.5) at all surface of the ends, $z = -L_c$ and $z = +L_c$, therefore, we are subjected to use the Saint-Venant's principle [8], which allows one to replace the so-called "weak" form of Eq.(B.8b) with the condition that the average normal stress, $\sigma_{zz}^{s,c}$, is equal to zero:

$$R_c^2 A_c + (R_s^2 - R_c^2)A_s = 0 \Leftrightarrow A_s\left(1 - \frac{R_s^2}{R_c^2}\right) = A_c. \tag{B.9a}$$

The solution of Eqs. (B.8a) is

$$C_s = R_s^2 B_s \text{ and } B_s\left(1 - \frac{R_s^2}{R_c^2}\right) = B_c. \tag{B.9b}$$

So that the in-plane stress components can be simplified as:

$$\sigma_{\rho\rho}^S = B_c \frac{1 - \frac{R_s^2}{\rho^2}}{1 - \frac{R_s^2}{R_c^2}}, \quad \sigma_{\psi\psi}^S = B_c \frac{1 + \frac{R_s^2}{\rho^2}}{1 - \frac{R_s^2}{R_c^2}}, \quad \sigma_{\rho\rho}^C = \sigma_{\psi\psi}^C = B_c. \tag{B.10}$$

Next, the conditions of displacement continuity can be rewritten as

$$\left[s_{11}\sigma_{\psi\psi}^S + s_{12}(\sigma_{zz}^S + \sigma_{\rho\rho}^S) - \{s_{11}\sigma_{\psi\psi}^C + s_{12}(\sigma_{zz}^C + \sigma_{\rho\rho}^C) + Q_{12}P_3^2\} + u_s - u_c\right]\Big|_{\rho = R_c} = 0, \tag{B.11a}$$

$$\left[s_{11}\sigma_{zz}^S + s_{12}(\sigma_{\psi\psi}^S + \sigma_{\rho\rho}^S) - \{s_{11}\sigma_{zz}^C + s_{12}(\sigma_{\psi\psi}^C + \sigma_{\rho\rho}^C) + Q_{11}P_3^2\} + u_s - u_c\right]\Big|_{\rho = R_c} = 0. \tag{B.11b}$$

The explicit form of Eqs.(B.11a-b) is

$$s_{12}A_s + \left(s_{11}\frac{1 + \frac{R_s^2}{R_c^2}}{1 - \frac{R_s^2}{R_c^2}} + s_{12}\right)B_c + u_s - u_c = s_{12}A_c + (s_{11} + s_{12})B_c + Q_{12}P_3^2, \tag{B.11c}$$

$$s_{11}A_s + \frac{2s_{12}B_c}{1 - \frac{R_s^2}{R_c^2}} + u_s - u_c = s_{11}A_c + 2s_{12}B_c + Q_{11}P_3^2. \tag{B.11d}$$

Taking into account (B.9a), one obtains:

$$s_{12}A_c + 2s_{11}B_c = \left(\frac{R_c^2}{R_s^2} - 1\right)(Q_{12}P_3^2 - u_s + u_c), \tag{B.11e}$$

$$s_{11}A_c + 2s_{12}B_c = \left(\frac{R_c^2}{R_s^2} - 1\right)(Q_{11}P_3^2 - u_s + u_c). \tag{B.11f}$$

The solution of the system of Eqs.(B.11e)- (B.11f) is

$$A_c = \left(\frac{R_c^2}{R_s^2} - 1\right)\left(\frac{s_{11}Q_{11} - s_{12}Q_{12}}{s_{11}^2 - s_{12}^2}P_3^2 - \frac{u_s - u_c}{s_{11} + s_{12}}\right), \tag{B.12a}$$

$$B_c = \frac{1}{2}\left(\frac{R_c^2}{R_s^2} - 1\right)\left(\frac{s_{11}Q_{12} - s_{12}Q_{11}}{s_{11}^2 - s_{12}^2}P_3^2 - \frac{u_s - u_c}{s_{11} + s_{12}}\right). \tag{B.12b}$$

Using Eq.(B.9), the rest of the constants are

$$A_s = \frac{R_s^2}{R_c^2}\left\{\frac{s_{11}Q_{11} - s_{12}Q_{12}}{s_{11}^2 - s_{12}^2}P_3^2 - \frac{u_s - u_c}{s_{11} + s_{12}}\right\}, B_s = \frac{R_c^2}{2R_s^2}\left\{\frac{s_{11}Q_{12} - s_{12}Q_{11}}{s_{11}^2 - s_{12}^2}P_3^2 - \frac{u_s - u_c}{s_{11} + s_{12}}\right\}. \tag{B.12c}$$

Using the constants (B.12) and expressions (B.6)-(B.7) for the components of the stresses and strains, we obtain the core srain as follows

$$u_{zz}^c = \frac{R_c^2}{R_s^2}\{u_c + Q_{11}P_3^2\} + \left(1 - \frac{R_c^2}{R_s^2}\right)u_s, \tag{B.13a}$$

$$u_{\rho\rho}^c = u_{\psi\psi}^c = \frac{R_c^2}{R_s^2}\{u_c + Q_{12}P_3^2\} + \left(1 - \frac{R_c^2}{R_s^2}\right)u_s + \left(1 - \frac{R_c^2}{R_s^2}\right)\frac{(s_{11} - s_{12})(u_c - u_s) + s_{11}Q_{12}P_3^2 - s_{12}Q_{11}P_3^2}{2(s_{11} + s_{12})}. \tag{B.13b}$$

Consideration of the surface tension in Eq.(B.8b) leads to the appearance of constant terms proportional to $-2s_{12}\frac{\mu}{R_s}$ and $-(s_{11} + s_{12})\frac{\mu}{R_s}$ in the right-hand side of expressions (B.13a) and (B.13b), respectively; but these terms are negligibly small (less than 0.01 %) and are set to zero for the considered radii of nanorods and nanowires ($R_s \geq 10$ nm) and realistic surface tension coefficient $\mu < 4$ N/m.

Finally, using Eq.(B.13), one could calculate the tetragonality $\frac{c_{lc}}{a_{lc}} \approx 1 + u_{zz}^c - u_{\rho\rho}^c$ as:

$$\frac{c_{lc}}{a_{lc}} = 1 + (Q_{11} - Q_{12})P_3^2 + \frac{1}{2}\left(\frac{R_c^2}{R_s^2} - 1\right)\left[\frac{\{(2s_{11}+s_{12})Q_{11} - (s_{11}+2s_{12})Q_{12}\}P_3^2}{s_{11}+s_{12}} + \frac{(s_{11}-s_{12})(u_c-u_s)}{s_{11}+s_{12}}\right] \quad (B.14)$$

Equation of state for polarization could be obtained by the minimization of the free energy (B.1) with respect to $P_3$, which yields:

$$2\{a_1 - Q_{11}\sigma_{zz}^c - Q_{12}(\sigma_{\rho\rho}^c + \sigma_{\psi\psi}^c)\}P_3 + 4a_{11}P_3^3 + 6a_{111}P_3^5 = E_3. \quad (B.15)$$

After substitution of the elastic stresses into Eq.(B.15) we obtain the equation of state with renormalized coefficients:

$$2\left\{a_1 - \left(1 - \frac{R_c^2}{R_s^2}\right)\frac{Q_{11}+Q_{12}}{s_{11}+s_{12}}(u_s - u_c)\right\}P_3 + 4\left\{a_{11} + \left(1 - \frac{R_c^2}{R_s^2}\right)\frac{s_{11}(Q_{11}^2+Q_{12}^2) - 2s_{12}Q_{11}Q_{12}}{2(s_{11}^2-s_{12}^2)}\right\}P_3^3 + 6a_{111}P_3^5 = E_3 \quad (B.16)$$

**APPENDIX C. Analytical calculations of the negative capacitance effect**

In the case $h_c \ll 2R_c$ the electric potential $\varphi$ of the three-layer capacitor, shown in **Fig. 8(a)** in the main text, is given by expressions:

$$\varphi_{s1}(x_3) = -\frac{x_3 - 2h_s - h_c}{\varepsilon_0 \varepsilon_s}D_3, \quad h_c + h_s \leq x_3 \leq h_c + 2h_s, \quad (C.1a)$$

$$\varphi_c(x_3) = \left(\frac{h_s + h_c - x_3}{\varepsilon_0 \varepsilon_b} + \frac{h_s}{\varepsilon_0 \varepsilon_s}\right)D_3 - \frac{1}{\varepsilon_0 \varepsilon_b}\int_{x_3}^{h_c+h_s} P(\tilde{z})d\tilde{z}, \quad h_s \leq x_3 \leq h_c + h_s, \quad (C.1b)$$

$$\varphi_{s2}(x_3) = -\frac{x_3}{\varepsilon_0 \varepsilon_s}D_3 + U, \quad 0 \leq x_3 \leq h_s. \quad (C.1c)$$

Here we introduce an electric displacement, $D_3 = \frac{\varepsilon_s h_c \bar{P} + \varepsilon_0 \varepsilon_s \varepsilon_b U}{2h_s \varepsilon_b + h_c \varepsilon_s}$, which is constant in all three layers. The electrode charge $Q$ of the three-layer capacitor is given by expression:

$$Q = -\varepsilon_0 \varepsilon_s \frac{d\varphi_{s1}}{dx_3}\bigg|_{x_3 = 2h_s + h_c} = \bar{P} \cdot \frac{h_c \varepsilon_s}{\varepsilon_s h_c + 2\varepsilon_b h_s} + \varepsilon_0 \frac{U \varepsilon_b \varepsilon_s}{\varepsilon_s h_c + 2\varepsilon_b h_s}. \quad (C.1d)$$

Here we introduce the average polarization $\bar{P}(x_1, x_2) = \frac{1}{h_c}\int_{h_s}^{h_s+h_c} P_3(x_1, x_2, \tilde{z})d\tilde{z}$.

The charge of the reference SrTiO$_3$ capacitor is $Q_r = C_r U$, where the reference capacitance is $C_r = \frac{\varepsilon_0 \varepsilon_s}{2h_s}$. The difference of the effective and reference capacitance is given by expression:

$$\Delta C = \frac{dQ}{dU} - C_r = \frac{d\bar{P}}{dU} \cdot \frac{h_c \varepsilon_s}{\varepsilon_s h_c + 2\varepsilon_b h_s} + \frac{\varepsilon_0 \varepsilon_b \varepsilon_s}{\varepsilon_s h_c + 2\varepsilon_b h_s} - \frac{\varepsilon_0 \varepsilon_s}{2h_s} \equiv \left(\frac{d\bar{P}}{dU} - \frac{\varepsilon_0 \varepsilon_s}{2h_s}\right)\frac{h_c \varepsilon_s}{\varepsilon_s h_c + 2\varepsilon_b h_s}. \quad (C.2)$$

The NC effect ($\Delta C > 0$) corresponds to the condition $\frac{d\bar{P}}{dU} > \frac{\varepsilon_0 \varepsilon_s}{2h_s}$. The magnitude of $\bar{P}$ can be estimated from the equation:

$$\left[\alpha_{SR} + \frac{2h_s}{\varepsilon_0(\varepsilon_s h_c + 2\varepsilon_b h_s)}\right]\bar{P} + \beta_R \overline{P_3^3} + \gamma \overline{P_3^5} + \delta \overline{P_3^7} = \frac{\varepsilon_s U}{\varepsilon_s h_c + 2\varepsilon_b h_s}, \quad (C.3)$$

where $\alpha_{SR} = 2\left(a_1 - \delta u\, \delta V \frac{Q_{11}+Q_{12}}{s_{11}+s_{12}}\right)$ (see Eq.(7a) in the main text). The average electric field corresponding to Eq.(C.3) is $E_3 = \frac{-\bar{P}}{\varepsilon_0} \frac{2h_s}{\varepsilon_s h_c + 2\varepsilon_b h_s} + \frac{\varepsilon_s U}{\varepsilon_s h_c + 2\varepsilon_b h_s}$. Note that the effective factor $\frac{2h_s}{\varepsilon_0(\varepsilon_s h_c + 2\varepsilon_b h_s)}$ in Eq.(C.3) coincides with the depolarization field factor $\frac{n_d}{\varepsilon_0[\varepsilon_b n_d + \varepsilon_s(1-n_d) + \varepsilon_s n_d(L_c/\lambda)]}$ in Eq.(2) in the particular case $\lambda \to h_s$, $L_c \to h_c/2$ and $n_d \to 1$ (which is a good approximation for $L_c/R_c \ll 1$).

Under the condition of negligibly small contribution of the nonlinear polarization powers in Eq.(C.3), the terms $\beta_R \overline{P_3^3} + \gamma \overline{P_3^5} + \delta \overline{P_3^7}$ can be omitted, and the derivative $\frac{d\bar{P}}{dU}$ can be estimated as:

$$\frac{d\bar{P}}{dU} \approx \frac{\varepsilon_s}{\varepsilon_s h_c + 2\varepsilon_b h_s}\left[\alpha_{SR} + \frac{2h_s}{\varepsilon_0(\varepsilon_s h_c + 2\varepsilon_b h_s)}\right]^{-1}. \quad (C.4)$$

Under the validity of Eq.(C.4), the NC effect can be reached under the conditions

$$\frac{\varepsilon_s}{\varepsilon_s h_c + 2\varepsilon_b h_s}\left[\alpha_{SR} + \frac{2h_s}{\varepsilon_0(\varepsilon_s h_c + 2\varepsilon_b h_s)}\right]^{-1} > \frac{\varepsilon_0 \varepsilon_s}{2h_s}, \qquad \alpha_{SR} + \frac{2h_s}{\varepsilon_0(\varepsilon_s h_c + 2\varepsilon_b h_s)} > 0. \quad (C.5)$$

The conditions (C.5) are equivalent to the condition:

$$-\frac{2h_s}{\varepsilon_0(\varepsilon_s h_c + 2\varepsilon_b h_s)} < \alpha_{SR} < 0. \quad (C.6)$$

The substitution of $\alpha_{SR} = 2\left\{a_1 - \delta u\, \delta V \frac{Q_{11}+Q_{12}}{s_{11}+s_{12}}\right\}$ in Eqs.(C.5)-(C.6) yields Eqs.(12) in the main text.